\documentclass[preprint,showpacs,preprintnumbers,amsmath,amssymb,nofootinbib,groupedaddress,
superscriptaddress,showkeys,notitilepage]{revtex4-1}

\usepackage[pdftex]{graphicx}
\usepackage{epstopdf}
\usepackage[bookmarks=true,bookmarksnumbered=true,bookmarkstype=toc, colorlinks=true, 
citecolor=blue, linkcolor=blue]{hyperref} 
\usepackage{units}

\usepackage{color}

\definecolor{pink}{cmyk}{0,0.5,0,0}

\usepackage{braket}

\usepackage{bm}

\usepackage{mathtools}% http://ctan.org/pkg/mathtools
\newcommand{\fsl}[1]{\ensuremath{\mathrlap{\!\not{\phantom{#1}}}#1}}% \fsl{<symbol>} Feynman Slash
\newcommand{\Slash}[1]{\ensuremath{\mathrlap{\!\not{\phantom{#1}}}#1}}% \fsl{<symbol>} Feynman Slash

\newcommand{\lsim}{\raise0.3ex\hbox{$\;<$\kern-0.75em\raise-1.1ex\hbox{$\sim\;$}}}
\newcommand{\gsim}{\raise0.3ex\hbox{$\;>$\kern-0.75em\raise-1.1ex\hbox{$\sim\;$}}}

\allowdisplaybreaks[4]

\newcommand{\lmlt}{L_\mu^{} - L_\tau^{}}
\newcommand{\mZp}{m_{Z'}}
\newcommand{\Am}{\overline{\mu}}
\newcommand{\At}{\overline{\tau}}

\usepackage{ulem}

\begin{document}

\title{Kinematical distributions of coherent neutrino trident production \\ in gauged $L_\mu-L_\tau$ model}

\author{Takashi Shimomura}
\email{shimomura@cc.miyazaki-u.ac.jp}
\affiliation{%
Faculty of Education, 
University of Miyazaki,
\\
1-1 Gakuen-Kibanadai-Nishi,
889-2192 Miyazaki,
Japan
}

\author{Yuichi Uesaka}
\email{uesaka@ip.kyusan-u.ac.jp}
\affiliation{%
Faculty of Science and Engineering, Kyushu Sangyo University, \\
2-3-1 Matsukadai, Higashi-ku, Fukuoka 813-8503, Japan
}

\date{\today}

\preprint{\bf UME-PP-015}

%%%%%%%%%%%%%%%%%%%%%%%%%%%%%%%%%%%%%%%%%%%%%%%%%%%%%%%%%%%%%%%%%%%%%%
 \begin{abstract}
We analyze the distributions of energy, opening angle and invariant mass in muonic neutrino 
trident production processes, $\nu_\mu \to \nu_\mu \mu \Am$, in a minimal gauged $U(1)_{\lmlt}$ model, 
in which the discrepancy of anomalous magnetic moment of muon can be solved. 
It is known that the total cross sections of the neutrino trident production are degenerate in new physics parameters, 
the new gauge coupling and gauge boson mass, and therefore other observables are needed to determine these 
parameters.  From numerical analyses, 
we find that the  muon energy and invariant mass distributions show the differences among the new physics 
parameter sets with which the total cross sections have the same value, while the anti-muon energy and opening angle distributions 
are not sensitive to the parameters.
\end{abstract}
%%%%%%%%%%%%%%%%%%%%%%%%%%%%%%%%%%%%%%%%%%%%%%%%%%%%%%%%%%%%%%%%%%%%%%

\maketitle

%%%%%%%%%%%%%%%%%%%%%%%%%%%%%%%%%%%%%%%%%%%%%%%%%%%%%%%%%%%%%%%%%%%%%%

%%%%%%%%%%%%%%%%%%%%%%%%%%%%%%%%%%%%%%%%%%%%%%%%%%%%%%%%%%
\section{Introduction}
%%%%%%%%%%%%%%%%%%%%%%%%%%%%%%%%%%%%%%%%%%%%%%%%%%%%%%%%%%
Anomalous magnetic moment of muon is a long-standing discrepancy between experimental measurements 
\cite{Bennett:2006fi, Tanabashi:2018oca} and theoretical predictions 
\cite{Blum:2018mom,Keshavarzi:2018mgv,Davier:2019can,Aoyama:2020ynm}.
The recent result of the Standard Model (SM) prediction \cite{Keshavarzi:2018mgv} shows that 
the difference of the anomalous magnetic moment, $a_\mu \equiv (g_\mu-2)/2$, from the measurements reaches to
\begin{align}
\Delta a_\mu \equiv a_\mu^\mathrm{Exp} - a_\mu^\mathrm{SM} = (27.06 \pm 7.26) \times 10^{-10}.
\end{align}
Thus the SM predictions is $3.7\sigma$ lower than the experimental measurements. 
Extensive studies on theoretical side have been made, however the discrepancy cannot be resolved within 
the SM of particle physics (for review see \cite{Lindner:2016bgg} for example). 
The E989 experiments at Fermilab \cite{Grange:2015fou} and the E34 experiment at J-PARC \cite{Abe:2019thb} are 
on-going and will reduce experimental uncertainties by a factor of four, which could confirm the discrepancy at $5\sigma$ level. 
Once the discrepancy is confirmed, it will be a clear signature of new physics (NP) beyond the SM.

Many new physics models have been proposed to explain the discrepancy of $a_\mu$ by extending the SM. 
One of the simplest extensions in this regard is to impose an extra $U(1)$ gauge symmetry on the SM, in which new 
contribution of a new gauge boson accounts for the deviation of the muon anomalous magnetic moment. 
Among such extensions, the $U(1)$ symmetry gauging flavor muon number minus tau flavor number or $\lmlt$ \cite{Foot:1990mn,He:1991qd,Foot:1994vd} has been gaining attention  in recent years.
 In \cite{Altmannshofer:2014pba}, it was shown that a gauge boson of the $U(1)_{\lmlt}$  symmetry can explain the deviation without conflicting experimental searches, provided that the mass and gauge coupling are  $\mathcal{O}(100)$ MeV 
and $\mathcal{O}(10^{-4})$, respectively. 
Possibilities on searches for this light and weakly interacting gauge boson have been studied in 
\cite{Gninenko:2014pea, Kaneta:2016uyt,Araki:2017wyg,Chen:2017cic,Nomura:2018yej,Banerjee:2018mnw,Jho:2019cxq,Iguro:2020rby,Amaral:2020tga}.  
Other studies based on the $\lmlt$ symmetry also have been done such as cosmic neutrino spectrum observed at IceCube \cite{Araki:2014ona,Araki:2015mya}, 
neutrino mass and mixing \cite{Asai:2017ryy,Asai:2018ocx,Asai:2019ciz,Araki:2019rmw}, 
dark matter \cite{Kamada:2018zxi, Gninenko:2018tlp, Foldenauer:2018zrz}, the baryon asymmetry of the Universe \cite{Asai:2020qax}, 
meson decay \cite{Ibe:2016dir,Han:2019diw,Jho:2020jsa} for recent works. 
Light gauge bosons interacting with muonic leptons can contribute to Neutrino Trident Production (NTP) processes such as $\nu_\mu + N \to \nu_\mu + \mu + \Am + N$ 
\cite{Czyz:1964zz, Lovseth:1971vv,Fujikawa:1971nx,Koike:1971tu,Koike:1971vg,Brown:1973ih,Belusevic:1987cw}. 
It was also shown in \cite{Altmannshofer:2014cfa,Altmannshofer:2014pba} that the NTP processes can set 
severe bound on  the gauge boson mass and the gauge coupling. 
Utilizing the results of the CHARM-II \cite{Geiregat:1990gz}, CCFR \cite{Mishra:1991bv} and 
NuTeV \cite{Adams:1999mn} experiments, one finds that the region of the mass above $\mathcal{O}(100)$ MeV and the gauge coupling 
above $\mathcal{O}(10^{-3})$ are excluded.
The analyses of the NTP processes in the SM or new physics models also have been done for future planned experiment, DUNE \cite{Magill:2016hgc,Magill:2017mps,Ballett:2018uuc,Altmannshofer:2019zhy,Ballett:2019xoj}, 
SHiP \cite{Magill:2016hgc,Magill:2017mps}, 
MINOS, No$\nu$A, MINERvA \cite{Ballett:2018uuc},
MicroBooNE \cite{deGouvea:2018cfv},
and on-going experiments, 
T2K \cite{Kaneta:2016uyt,Ballett:2018uuc}, 
IceCube \cite{Ge:2017poy,Zhou:2019vxt,Beacom:2019pzs} 
taking into account coherent and diffractive processes. 
In particular, the liquid argon detector at the near site in the DUNE experiment is expected to observe $\mathcal{O}(100)$ events of muonic 
NTP process \cite{Ballett:2018uuc,Altmannshofer:2019zhy,Ballett:2019xoj}.
As presented in these works, the contours of the total cross section of the NTP processes are obtained as 
a function of new physics parameters, i.e. the mass and coupling constant of new gauge bosons. 
This fact results in that the new physics parameters cannot be determined uniquely only by measurements of 
the total cross sections. In other words, the total cross sections are degenerate in the new physics parameters. 
To determine or further constrain the new physics parameters, one needs other observables in addition to  the total cross sections. 
One of such observables will be the differential cross sections that are generally measured simultaneously 
in experiments. 
When the differential cross sections show the differences to the new physics parameters for the fixed values of the 
total cross section, we can determine or constrain the parameters by combining the information from the differential and total cross sections.
As a first step for this purpose, we analyze the parameter dependences of 
the differential cross sections with respect to the energies, opening angle and invariant masses of the final state muons 
in a minimal $\lmlt$ model. 
Our results will show which distributions should be used for detailed analyses for the determination of the parameters.

This paper is organized as follows. In Sec.~II, we briefly review a minimal gauged $\lmlt$ model and 
present relevant interactions. The amplitudes and cross section of NTP processes are 
given in Sec.~III. Then, we show our numerical results on the distributions with respect to the energy, opening angle and invariant mass of muon pair in Sec.~IV. Section V is devoted to summary.

%%%%%%%%%%%%%%%%%%%%%%%%%%%%%%%%%%%%%%%%%%%%%%%%%%%%%%%%%%
\section{Minimal $L_\mu - L_\tau$ Model} \label{sec:model}
%%%%%%%%%%%%%%%%%%%%%%%%%%%%%%%%%%%%%%%%%%%%%%%%%%%%%%%%%%
%
\begin{table}[t]
  \begin{center}
    \begin{tabular}{|c|c|c|c|c|c|c|}\hline
      &
      $~~e~~$ & 
      $~~\mu~~$ & 
      $~~\tau~~$ & 
      $~~\nu_e~~$ & $~~\nu_\mu~~$ & $~~\nu_\tau~~$ \\ \hline
      $~~U(1)_{L_\mu -L_\tau}~~$ & $0$ & $1$ & $-1$ & $0$ &$1$ & $-1$ \\ \hline
    \end{tabular}
  \end{center}
  \caption{The charge assignment of the gauged $U(1)_{L_\mu - L_\tau}$ model. }
  \label{charge}
\end{table}
We start our discussion with reviewing a minimal gauged $\lmlt$ model. 
The gauge sector of the SM is extended by adding the $U(1)_{L_\mu - L_\tau}$ gauge symmetry under which 
mu and tau flavored leptons among the SM fermions are charged. 
The charge assignment for leptons under this symmetry is shown in Table \ref{charge}. 
In the table, $e,~\mu$ and $\tau$ represent 
charged leptons, and $\nu_e$, $\nu_\mu$ and $\nu_\tau$ are corresponding left-handed neutrinos, respectively. 
Up-type and down-type quarks as well as the Higgs boson are singlet 
under the $U(1)_{\lmlt}$ gauge symmetry.

The relevant interaction Lagrangian for the NTP processes is given by
\begin{align}
\mathcal{L}_{\mathrm{int}} &= eA_\rho J^\rho_{\mathrm{em}} 
	-\frac{4G_F}{\sqrt{2}} 
	[\overline{\nu_{\ell_4}} \gamma_\rho \nu_{\ell_3}] 
	[\overline{\ell}_2 \gamma^\rho (g_L P_L + g_R P_R) \ell_1]
        + g' Z'_\rho J_{Z'}^\rho  \label{eq:lagrangian} 
\end{align}
where $e$, $A^\rho$ and $J^\rho_{\mathrm{em}}$ are the elementary electric charge, photon field and 
electromagnetic current of the SM, respectively.
In the second term of Eq.~\eqref{eq:lagrangian}, 
$G_F$ is the Fermi coupling constant, and $\ell$ and $\nu_\ell$ are a charge lepton and a neutrino with flavor $\ell_i = e, \mu, \tau~(i=1\mathrm{-}4)$.
The left-handed (right-handed) projection operator is denoted as $P_{L(R)}$.  
The constants $g_L$ and $g_R$ are given by 
\begin{subequations}
\begin{align}
g_L &= \left(-\frac{1}{2} + \sin^2 \theta_W\right)\delta_{\ell_1,\ell_2}\delta_{\ell_3,\ell_4}+\delta_{\ell_1,\ell_4}\delta_{\ell_2,\ell_3}, \label{eq:gL}\\
g_R &= \sin^2 \theta_W\delta_{\ell_1,\ell_2}\delta_{\ell_3,\ell_4}, \label{eq:gR}
\end{align}
\label{eq:gLR}
\end{subequations}
where $\theta_W$ is the Weinberg angle. 
From Eq.~\eqref{eq:gL}, $g_L$ for muonic ($\nu_\mu \to \nu_\mu \mu \Am$) and tauonic ($\nu_\mu \to \nu_\mu \tau \At$) NTP processes is
\begin{subequations}
\begin{align}
g_L = \begin{cases}
\frac{1}{2} + \sin^2\theta_W~~~~~(\nu_\mu \to \nu_\mu \mu \Am), \\
-\frac{1}{2} + \sin^2\theta_W~~~(\nu_\mu \to \nu_\mu \tau \At),
\end{cases}
\end{align}
\end{subequations}
respectively, while from Eq.~\eqref{eq:gR}, $g_R$ is $\sin^2\theta_W$ for both processes.
The third term of Eq.~\eqref{eq:lagrangian} is the interaction of the $\lmlt$ gauge boson $Z'$  
with the gauge coupling constant $g'$. 
The $\lmlt$ gauge current, $J^\rho_{Z'}$, is  given by
\begin{align}
J_{Z'}^\rho &=
 \overline{\mu} \gamma^\rho \mu 
 -\overline{\tau} \gamma^\rho \tau
+\overline{\nu_\mu} \gamma^\rho \nu_\mu
 -\overline{\nu_\tau} \gamma^\rho \nu_\tau. \label{eq:current}
\end{align}
%

%%%%%%%%%%%%%%%%%%%%%%%%%%%%%%%%%%%%%%%
\begin{figure}[t]
\begin{tabular}{ccc}
	\includegraphics[width=55mm,clip]{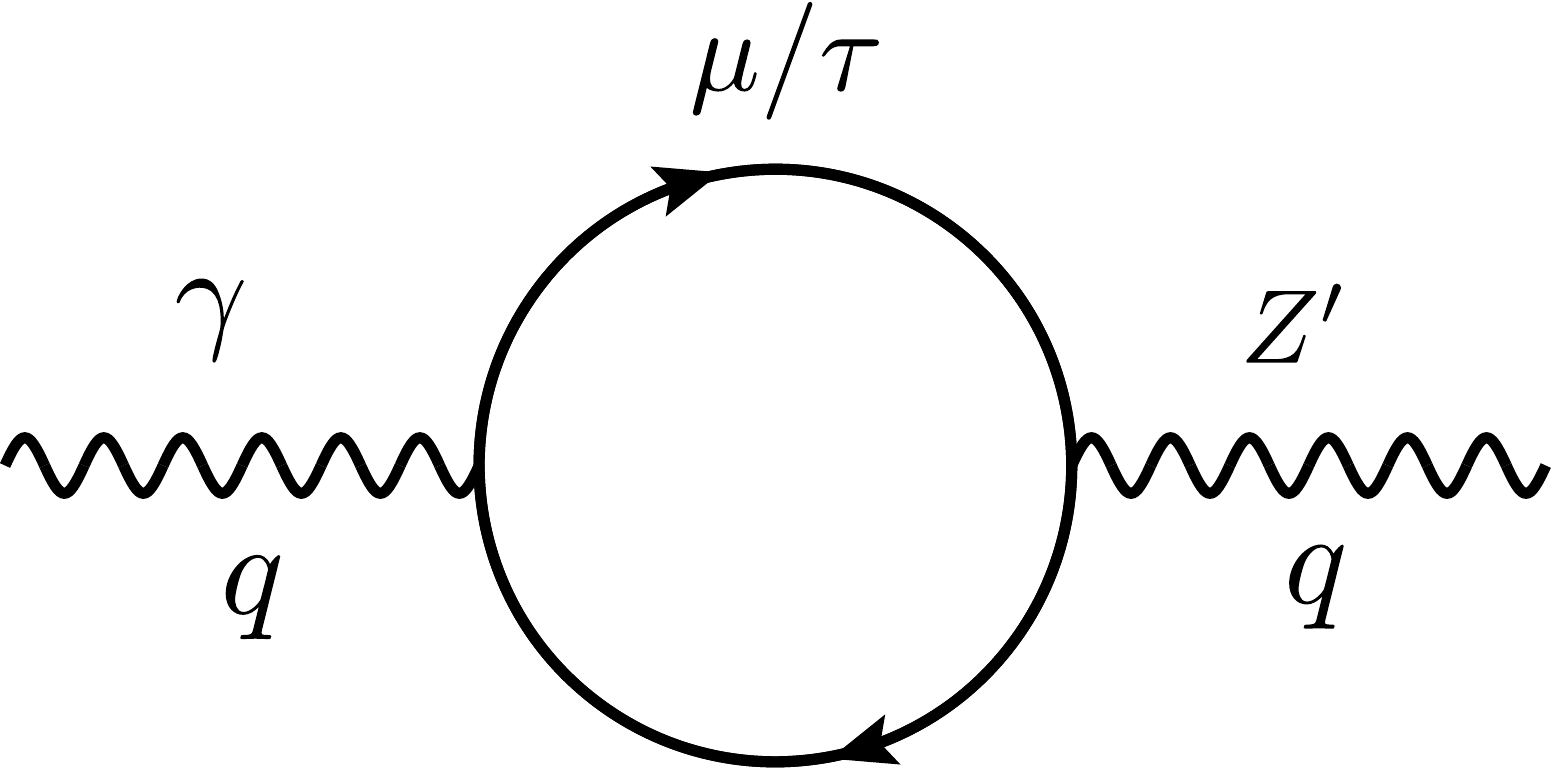}\\
\end{tabular}
\caption{Loop induced kinetic mixing between photon $\gamma$ and the $Z'$ boson.}
\label{fig:kinetic-mixing}
\end{figure}
%%%%%%%%%%%%%%%%%%%%%%%%%%%%%%%%%%%%%%%
In this work, we consider a minimal $\lmlt$ model in which the gauge kinetic mixing term between 
the $U(1)_Y$ hypercharge and  $U(1)_{\lmlt}$ symmetries is absent at tree-level.
Even though, the gauge kinetic mixing can be generated radiatively via loop diagrams in which muon, tau and 
neutrinos propagate. The loop-induced kinetic mixing parameter between photon $\gamma$ and $Z'$ can be 
obtained at one-loop level by 
evaluating Fig.\ref{fig:kinetic-mixing} as
\begin{align}
\epsilon(q^2) = \frac{8 e g'}{(4 \pi)^2} \int^1_0 dx x(1-x) 
\log\left( \frac{m_\tau^2 - x(1-x) q^2}{m_\mu^2 - x(1-x) q^2}  \right)
, \label{eq:loop-induced-mix}
\end{align}
where $q$ is the four momentum carried by $\gamma$ and $Z'$, and $m_\mu$ and $m_\tau$ are the mass 
of muon and tau, respectively.  The approximate expression of Eq.~\eqref{eq:loop-induced-mix} is given by
\begin{align}
\epsilon(q^2) \simeq 
\begin{cases}
 \frac{8 eg'}{3 (4 \pi)^2} \log \frac{m_\tau}{m_\mu},~~~(q^2 \ll 4m_\mu^2), \\
-\frac{6 eg'}{(4 \pi)^2} \left\{ 
\left( \frac{m_\tau^2}{q^2} - \frac{m_\mu^2}{q^2} \right)
+ i \pi \left( \frac{m_\tau^4}{q^4} - \frac{m_\mu^4}{q^4} \right)
\right\},~~~(q^2 \gg 4 m_\tau^2).
\end{cases}
\end{align}
This loop-induced kinetic mixing parameter is about two orders of 
magnitude smaller than $g'$ for $q^2 \ll 4 m_\mu^2$. It is further suppressed by a power of $m_{\tau,\mu}^2/q^2$ 
for $q^2 \gg 4 m_\tau^2$. For the intermediate $q^2$ $(4 m_\mu^2 < q^2 < 4 m_\tau^2)$, the real and imaginary parts 
are also two orders of magnitude smaller than $g'$. Therefore it is negligible compared with $g'$. We drop the 
loop-induced kinetic mixing parameter in our analyses.  
There also exists the loop-induced kinetic mixing between $Z'$ and the neutral weak boson $Z$.  However, 
since the energy of incident neutrinos we consider is smaller than the $Z$ boson mass, $m_Z$, such a mixing 
is practically negligible because it is suppressed by $m_Z$.

We also assume that the $\lmlt$ symmetry as well as the EW symmetry are appropriately broken without 
conflicting all existing experimental data so that $Z'$ can acquire a mass $\mZp$ of order $0.01-10$ GeV. 
We do not specify the scalar sector of the model and treat $m_{Z'}$ as a free 
parameter in the following analyses.
Thus only two parameters, $\mZp$ and $g'$, are newly introduced to the SM in our setup.

%%%%%%%%%%%%%%%%%%%%%%%%%%%%%%%%%%%%%%%%%%%%%%%%%%%%%%%%%%
\section{Neutrino Trident Production Processes}
%%%%%%%%%%%%%%%%%%%%%%%%%%%%%%%%%%%%%%%%%%%%%%%%%%%%%%%%%%
In this section, the amplitudes and cross sections of the NTP processes in the SM and 
the minimal gauged $U(1)_{\lmlt}$ model are presented, as well as a brief summary of the experimental results.
Depending on the virtuality of the photon $q^2$, the appropriate picture of a hadronic target is different.
Based on the appropriate hadronic picture, the NTP can be classified to three processes: coherent, diffractive, and deep inelastic, where the incoming neutrino scatters off the nuclei, nucleons, and quarks, respectively.
According to Ref.~\cite{Magill:2016hgc}, the deep inelastic contribution accounts for at most 1\% of the total NTP cross section, and therefore we do not consider this contribution.
For relevant energies of the initial neutrino, the coherent and diffractive processes give comparable contributions.
As the first step, we focus on the coherent process in this work.

\begin{figure}[t]
\begin{tabular}{ccc}
	\includegraphics[width=120mm,clip]{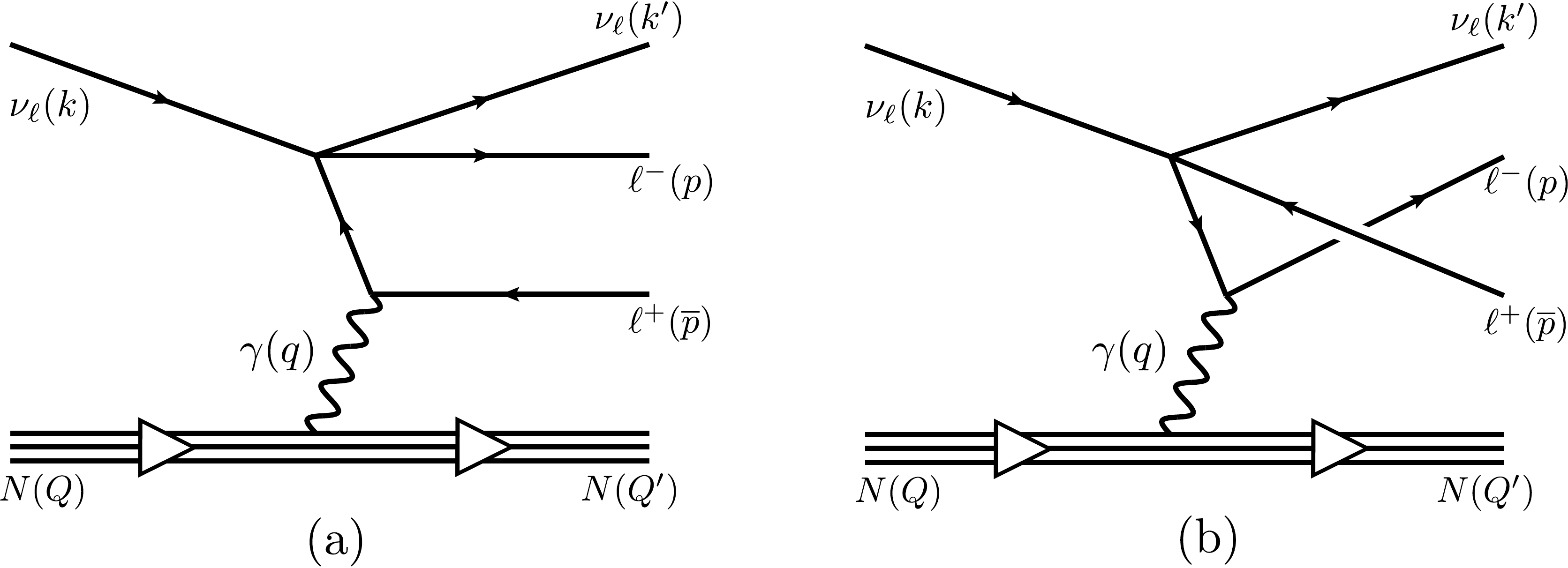}\\
\end{tabular}
\caption{Feynman diagrams of the NTP processes in the SM and four momentum assignment.}
\label{fig:trident_picture}
\end{figure}

In the following subsections, the four momenta of incident neutrino $(\nu_\ell)$ and nucleus $(N)$ are assigned to $k$ and $Q$ 
while those of outgoing ones are assigned to $k'$ and $Q'$, respectively. 
For lepton $(\ell^-)$ and anti-lepton $(\ell^+)$, the four momenta are assigned to $p$ and $\overline{p}$, and 
for virtual photon, the momentum is denoted as $q$. 
The Feynman diagrams of the NTP processes in the SM are shown in Fig.\ref{fig:trident_picture}.

%%%%%%%%%%%%%%%%%%%%%%%%%%%%%%%%%%%%%%%
%%%%%%%%%%%%%%%%%%%%%%%%%%%%%%%%%%%%%%%

%%%%%%%%%%%%%%%%%%%%%%%%%%%%%%%%%%%%%%%%%%%%%%%%%%%%%%%%%%
\subsection{Experimental Results}
%%%%%%%%%%%%%%%%%%%%%%%%%%%%%%%%%%%%%%%%%%%%%%%%%%%%%%%%%%
The muonic NTPs, $\nu_\mu \to \nu_\mu \mu \Am$, has been measured by the CHARM-II \cite{Geiregat:1990gz}, CCFR \cite{Mishra:1991bv} and NuTeV \cite{Adams:1999mn} experiments. The results are given as the ratio of the observed cross 
section to the SM prediction, $\sigma_\mathrm{SM}$,  
\begin{subequations}
\begin{align}
\frac{\sigma_\mathrm{CHARM-II}}{\sigma_\mathrm{SM}} &= 1.58 \pm 0.57, \\
\frac{\sigma_\mathrm{CCFR}}{\sigma_\mathrm{SM}} &= 0.82 \pm 0.28, \\
\frac{\sigma_\mathrm{NuTeV}}{\sigma_\mathrm{SM}} &= 0.72^{+1.73}_{-0.72}.
\end{align}
\end{subequations}
The CHARM-II and CCFR results are 
consistent with the SM prediction within the error. The NuTeV result has relatively large uncertainty and includes null result.
Therefore we use the CHARM-II and CCFR results for our analyses.

%%%%%%%%%%%%%%%%%%%%%%%%%%%%%%%%%%%%%%%%%%%%%%%%%%%%%%%%%%
\subsection{Amplitudes}
%%%%%%%%%%%%%%%%%%%%%%%%%%%%%%%%%%%%%%%%%%%%%%%%%%%%%%%%%%
From Eq.~\eqref{eq:lagrangian}, the SM amplitude of the NTP processes in Fig.~\ref{fig:trident_picture} is given by
\begin{align}
\mathcal{M}_{\mathrm{SM}} = \frac{4 e^2 G_F}{\sqrt{2}} 
\big[ \overline{u_\nu}(k') \gamma^\alpha P_L u_\nu(k) \big]
\big[ \overline{u_\ell}(p)  O^\mu_\alpha v_\ell(\overline{p}) \big]
\frac{1}{q^2} \braket{Q' | J_\mu(-q^2) | Q}, \label{eq:amp-sm}
\end{align}
where $u_\ell~(v_\ell)$ and $u_\nu$ are the spinor of charged (anti-)lepton and neutrino, respectively. 
The operator $O^\mu_\alpha$ 
represents the charged lepton current part which is defined by
\begin{align}
O^\mu_\alpha = 
\gamma^\mu \frac{\fsl{p} + \fsl{q} + m_\ell}{(p+q)^2 - m_\ell^2} \gamma_\alpha (g_L P_L + g_R P_R) 
+ \gamma_\alpha (g_L P_L + g_R P_R) \frac{-\fsl{\overline{p}} - \fsl{q} + m_\ell}{(\overline{p} + q)^2 - m_\ell^2} 
\gamma^\mu. \label{eq:amp-o}
\end{align}
where $m_l$ is the mass of the charged lepton. Throughout this paper, neutrinos are assumed to be massless.
Note that $O^\mu_\alpha$ satisfies the current conservation condition,
$q_\mu O^\mu_\alpha = 0$ \cite{Fujikawa:1971nx}.
The operator $J_\mu$ in the braket product is the electromagnetic current for nucleus.

From Eqs.~\eqref{eq:amp-sm} and \eqref{eq:amp-o}, the squared amplitude with summing  over spins is obtained as
\begin{align}
\sum_{\mathrm{spins}} |\mathcal{M}_{\mathrm{SM}}|^2 = \frac{e^4 G_F^2}{2 q^4} 
j^{\alpha \beta} L_{\alpha \beta}^{\mu \nu} J_{\mu \nu} ,
\label{eq:ampsq-sm}
\end{align}
where $j^{\alpha\beta}$, $L^{\mu\nu}_{\alpha\beta}$ and $J_{\mu \nu}$ represent neutrino, charged lepton and nucleus contributions, respectively. These tensors are defined as
\begin{subequations}
\begin{align}
j^{\alpha \beta} &= 8 (k^\alpha {k'}^\beta + k^\beta {k'}^\alpha - k \cdot k' g^{\alpha \beta} 
- i \epsilon^{\rho \alpha \sigma \beta} k_\rho k'_\sigma), \label{seq:neutrino-part}\\
L^{\mu \nu}_{\alpha \beta} &= 4 \mathrm{Tr}
	\big[ (\fsl{p} + m_\ell) W^\mu_\alpha (g_L P_L + g_R P_R) (\fsl{\overline{p}} - m_\ell ) V^\nu_\beta 
		(g^\ast_L P_L + g^\ast_R P_R) \big], \label{seq:charged-part} \\
J_{\mu \nu} &= \braket{Q' | J_\mu(t) | Q} \braket{Q | J_\nu^\dagger(t) |Q'}. \label{seq:nuclear-part}
\end{align}
\label{eq:ampsq-parts}
\end{subequations}
where $t$ is defined by $t = -q^2 = -(Q'-Q)^2$.
For convenience, we express the contraction of three tensors as 
\begin{align}
j^{\alpha\beta}L_{\alpha\beta}^{\mu\nu}J_{\mu\nu} = |g_L|^2 M_L + |g_R|^2 M_R 
- (g_L g_R^\ast  + g_L^\ast g_R) M_{LR}, \label{eq:lepton-part}
\end{align}
where the concrete forms of $M_L,~M_R$ and $M_{LR}$ 
are given in Appendix \ref{apdx:clep-tensor}. 
Since the cross term of $g_L$ and $g_R$, $M_{LR}$, is proportional to $m_\ell^2$, it is sub-leading when the lepton mass $m_\ell$ is small compared to the energy scale of the NTP process.
Each term in Eq.~\eqref{eq:lepton-part} is invariant under the exchange of the lepton momenta, $(p,k) \leftrightarrow (\overline{p},k')$. 
Furthermore, under the exchange of either $p \leftrightarrow \overline{p}$ or $k \leftrightarrow k'$, 
$M_L$ and $M_R$ are exchanged each other,
\begin{align}
M_L^{\mu \nu} \longleftrightarrow M_R^{\mu \nu}, \label{eq:ML-MR-exchange}
\end{align}
while $M_{LR}$ remains the same.
The nucleus tensor, Eq.~\eqref{seq:nuclear-part},  can be expressed in terms of a nuclear form factor. 
For spin-$0$ nucleus, 
\begin{align}
J_{\mu \nu} = Z^2 (Q + Q')_\mu (Q+Q')_\nu |F(t)|^2, \label{eq:Jmunu}
\end{align}
where $Z$ is the atomic number of nucleus, and $F(t)$ is the nuclear form factor given by 
\begin{align}
F(t) &=4\pi\int_{0}^{\infty}drr^2\rho(r)\frac{\sin\sqrt{t}r}{\sqrt{t}r}. 
\end{align}
and $\rho(r)$ is the nuclear density.
The normalization condition for $\rho(r)$ is
\begin{align}
4\pi\int_{0}^{\infty}drr^2\rho(r)=1,
\end{align}
and the integral variable $r$ is a distance from the center of nucleus.

According to Ref.~\cite{Ballett:2018uuc,Altmannshofer:2019zhy,Ballett:2019xoj}, the DUNE experiment will provide us a large number of NTP events, where liquid argon is used at the near detector.
In our numerical analysis, we consider argon as the target nucleus.
Following Ref.~\cite{DeJager:1974liz}, we parametrize $\rho$ as
\begin{align}
\rho(r)=\rho_0\frac{1+w\dfrac{r^2}{c^2}}{1+\exp\left(\dfrac{r-c}{z}\right)}, \label{eq:nuc-density}
\end{align}
where $\rho_0$ is a normalization factor. 
The parameters are given as $c=3.73$fm, $z=0.62$fm, $w=-0.19$ for $^{40}$Ar.

%%%%%%%%%%%%%%%%%%%%%%%%%%%%%%%%%%%%%%%
\begin{figure}[t]
\begin{tabular}{ccc}
	\includegraphics[width=80mm,clip]{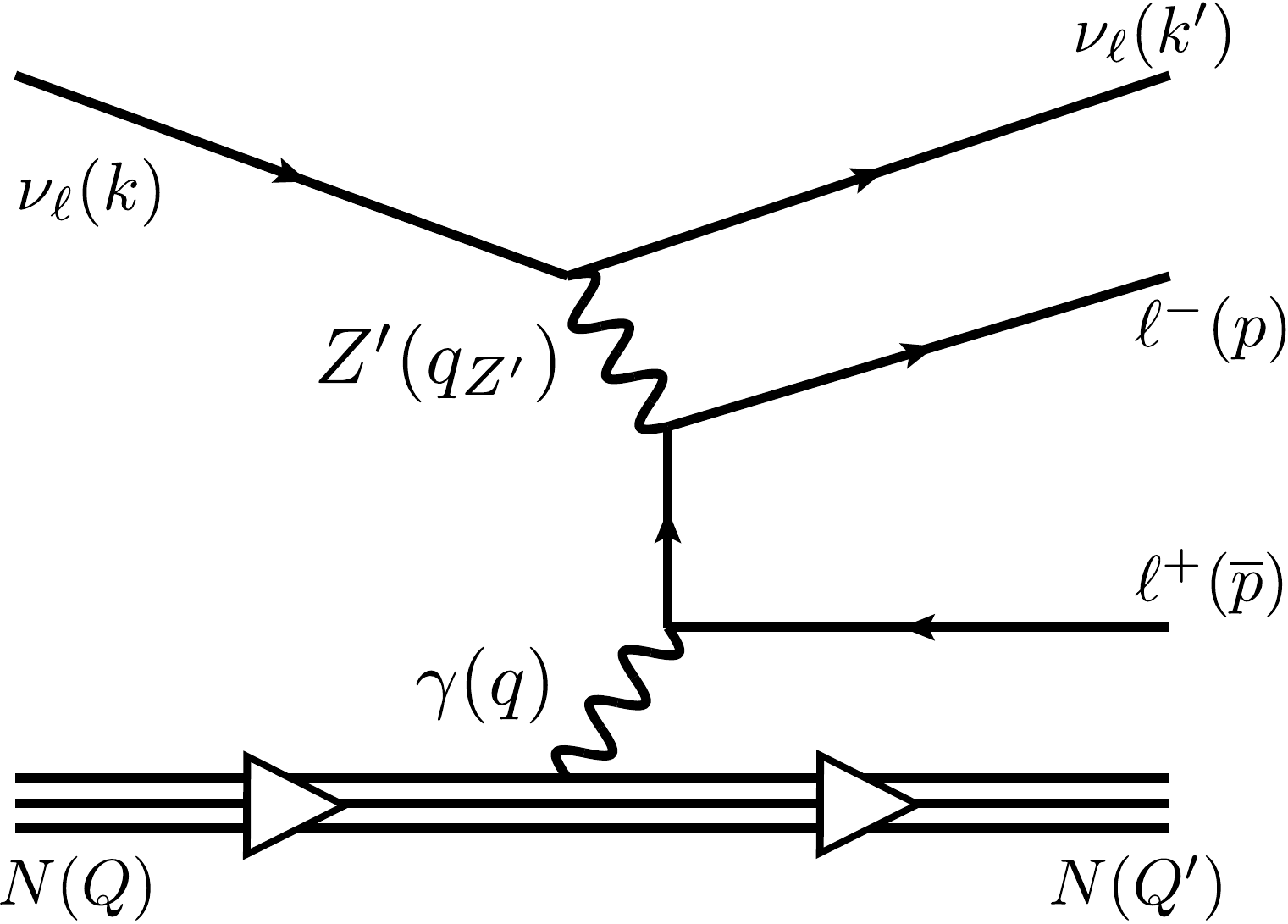}\\
\end{tabular}
\caption{The $Z'$ contribution of Feynman diagrams of NTP.}
\label{fig:trident_picture_03}
\end{figure}
%%%%%%%%%%%%%%%%%%%%%%%%%%%%%%%%%%%%%%%
The $Z'$ contribution to the NTP processes is shown in Fig.~\ref{fig:trident_picture_03}. 
The amplitude has the same spinor structure with Eq.~\eqref{eq:amp-sm}. 
Only difference between the SM amplitude and the $Z'$ amplitude is the propagator of $Z'$ instead of $G_F$. 
Thus, the total amplitude squared of the NTP processes in our model, $|\mathcal{M}_{\mathrm{total}}|^2$, is obtained by 
simply replacing $g_L$ and $g_R$ in Eq.~\eqref{seq:charged-part} as 
\begin{align}
g_{L(R)} \to g_{L(R)} \mp \frac{\sqrt{2}}{4 G_F} \frac{g'^2}{q_{Z'}^2 - \mZp^2}, \label{eq:coupling}
\end{align}
for $\nu_\mu \to \nu_\mu \mu \Am$ ($-$) and $\nu_\mu \to \nu_\mu \tau \At$ ($+$). 
From Eq.~\eqref{eq:coupling}, the NP parameter dependence disappears when 
\begin{align}
|q_{Z'}^2| + \mZp^2 \gg \frac{\sqrt{2}}{4 G_F} {g'}^2 \simeq (157~\mathrm{MeV})^2 \left( \frac{g'}{9 \times 10^{-4}} \right)^2.
\label{eq:ntp-cond}
\end{align}
As we will see in the next section, the $Z'$ contribution is negligible in the tauonic NTP process. This fact suggests that 
$|q_{Z'}^2|$ will become larger as the final state leptons are heavier.
Then, the cross section is almost the same as that of the SM for the tauonic NTP process.
In the next subsection, we show  the dependence of the total cross section on 
the new physics parameter $g'$ and $\mZp$.

%%%%%%%%%%%%%%%%%%%%%%%%%%%%%%%%%%%%%%%%%%%%%%%%%%%%%%%%%%
\subsection{Trident Production Cross Section} \label{subsec:ntp-cs}
%%%%%%%%%%%%%%%%%%%%%%%%%%%%%%%%%%%%%%%%%%%%%%%%%%%%%%%%%%
The total cross section of the NTP is given by 
\begin{align}
\sigma = \frac{1}{2(s-M^2)}\int d\Pi \sum_{\mathrm{spins}} |\mathcal{M}_{\mathrm{total}}|^2,
\end{align}
where $s = (k+Q)^2$ is the center of mass energy and $d\Pi$ is the phase space integral measure given by
\begin{align}
d\Pi = \frac{d^3 k'}{(2 \pi)^3 2 E_{k'}} \frac{d^3 p}{(2 \pi)^3 2 E_p} \frac{d^3 \overline{p}}{(2 \pi)^3 2 E_{\overline{p}}}
\frac{d^3 Q'}{(2 \pi)^3 2 E_{Q'}} (2 \pi)^4 \delta^{(4)} (k' + p + \overline{p} + Q' - k -Q). 
\end{align}
By the energy-momentum conservations and rotational symmetry, the number of the integrals can be reduced 
from twelve to seven. Then, we perform the phase space integrations numerically 
using the changes of the integral variables shown in Appendix \ref{apdx:phase-space-int}.

\begin{figure}[t]
\begin{tabular}{ccc}
	\includegraphics[width=0.49\linewidth]{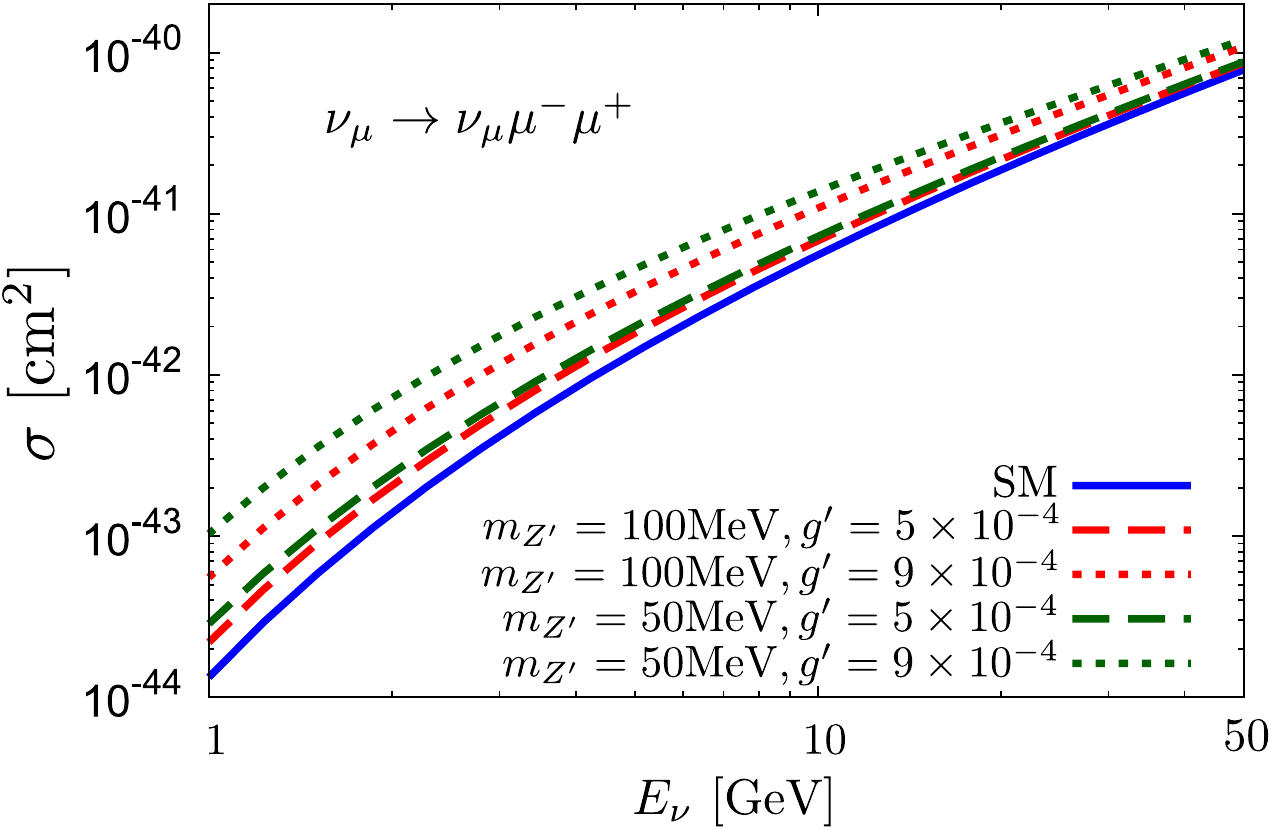} &
	~ &
	\includegraphics[width=0.49\linewidth]{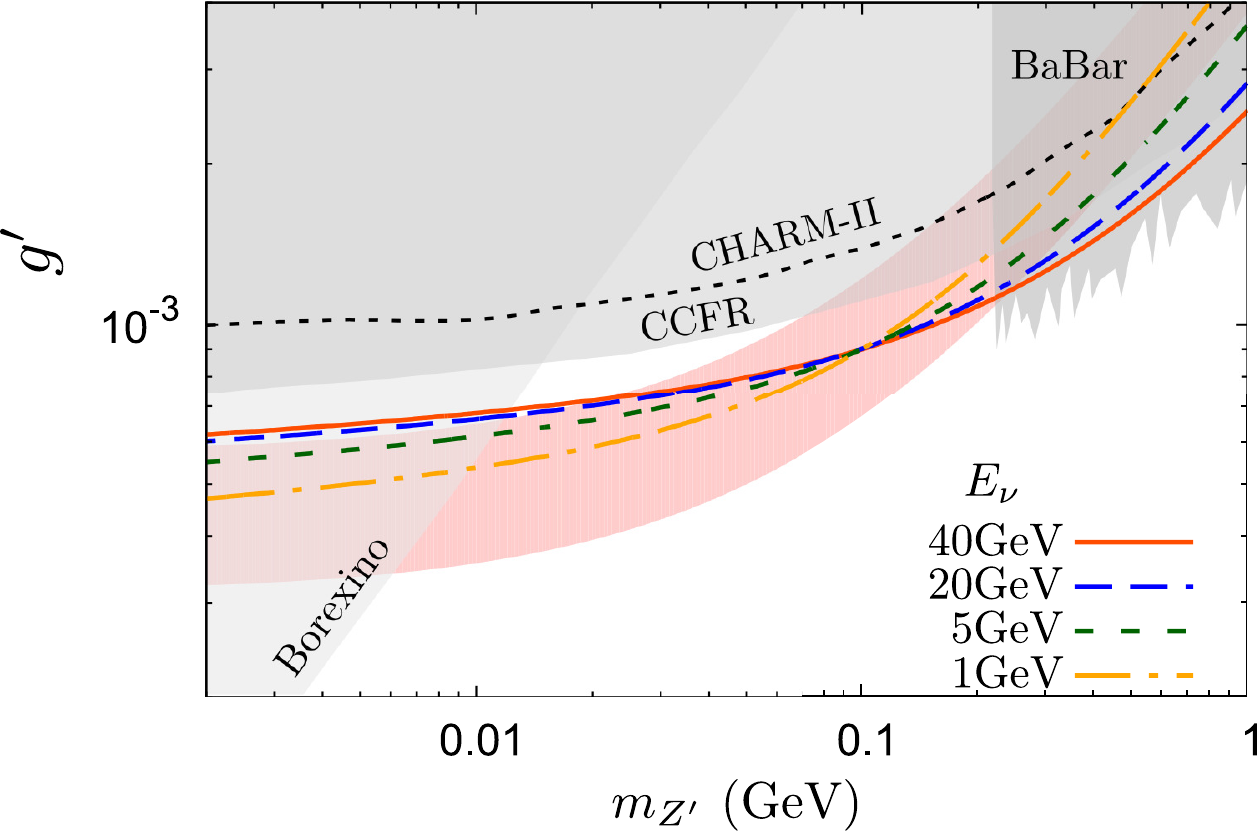} 
\end{tabular}
\caption{Left: Incident neutrino beam energy dependence of $\nu_\mu \to \mu_\mu \mu \Am$ cross section  
in the $L_\mu - L_\tau$ model. 
Right: Contour plots of the same cross section for $\sigma = 10^{-40}$ cm$^2$  in $m_{Z'}$-$g'$ plane.
}
\label{fig:total-cs}
\end{figure}

The cross sections of muonic NTP in the minimal $L_\mu - L_\tau$ model 
are shown in Fig.~\ref{fig:total-cs}. 
The left panel shows the cross sections with $m_{Z'} = 50, ~100$ MeV and $g' = 5 \times 10^{-4},~9\times 10^{-4}$ as a function of the incident neutrino energy $E_\nu$.
The right panel shows the contours of the muonic NTP cross section in our model.
In the left panel, the parameters, $(\mZp, g')$, are taken from the right panel 
as illustrating examples. 
Red curves correspond to $\mZp = 100$ MeV and green ones to $50$ MeV, respectively. 
For comparison, we also show the cross section in the SM with the blue solid curve. 
Our results of the total cross section in the SM are in good agreement with previous studies \cite{Brown:1973ih,Belusevic:1987cw} and 
\cite{Magill:2016hgc,Ballett:2018uuc}. 
It can be seen that the NP contributions become smaller compared with the SM cross section as 
$E_\nu$ becomes higher. This behavior generally holds even for $Z'$ with much lighter mass than $E_\nu$. 
As we explained in the previous subsection, 
this is because $\left|q_{Z'}^2\right|$ can take larger values than $\mZp^2$ for higher $E_\nu$, the NP contribution or the propagator of $Z'$ 
decreases as $g'^2/(G_F q_{Z'}^2)$. 
Thus, the cross section is less sensitive to the NP parameters for higher $E_\nu$, which has been shown in Ref.~\cite{Kaneta:2016uyt}.
Higher resolutions on momentum and/or energy measurements are required to solve the 
degeneracy in higher beam experiments like DUNE. 
On the other hand, for smaller $E_\nu$, the NP contributions to the cross section becomes larger, 
but the cross section itself becomes smaller. For example, for $E_\nu = 1$ GeV, the cross section 
is $\mathcal{O}(10^{-44})$ cm$^2$. 
%

%%%%%%%%%%%%%%%%%%%%%%%%%%%%%%%%%%%%%%%
%%%%%%%%%%%%%%%%%%%%%%%%%%%%%%%%%%%%%%%
In the right panel of Fig.~\ref{fig:total-cs}, red, blue, green and orange curves correspond to the same cross sections for $E_\nu = 40,~20,~5$ and $1$ GeV, respectively.
We chose $(m_Z',g')=(0.1\,\text{GeV},9\times 10^{-4})$ as a reference parameter set to determine the values of the cross section. 
Thus all curves intersect at this point.
The pink band represents muon $g-2$ favored region within $2\sigma$ and the gray shaded regions are excluded by 
Borexino \cite{Kaneta:2016uyt}\footnote{The constraints from Borexino are discussed 
in \cite{Harnik:2012ni,Agarwalla:2012wf,Bilmis:2015lja} in various different scenarios of new force. The constraint is translated from a $B-L$ gauge symmetric model in \cite{Kaneta:2016uyt}. }, 
CHARM-II \cite{Geiregat:1990gz}, CCFR \cite{Mishra:1991bv} and BaBar \cite{TheBABAR:2016rlg}. 
This plot clearly shows that the cross section is degenerate in $\mZp$ and $g'$ over wide range. 
As we mentioned in the introduction, for the determination of the parameters, one needs additional information besides the cross section value.

%%%%%%%%%%%%%%%%%%%%%%%%%%%%%%%%
\begin{table}[t]
\begin{tabular}{|c|c|c|c||c|c|c|c|} \hline
~~$E_\nu$~~ & ~~~ $m_{Z'}$ ~~~ & ~~~~~~ $g'$ ~~~~~~ & ~~~~~~~~ $\sigma$ ~~~~~~~~ &
~~$E_\nu$~~ & ~~~ $m_{Z'}$ ~~~ & ~~~~~~ $g'$ ~~~~~~ & ~~~~~ $\sigma$ ~~~~~ \\ \hline \hline
$1$     & ---           & ---                                  & $1.33 \times 10^{-3}$   & $20$    & ---          & ---                                 & $1.8945$ \\ \hline
           & $0.020$ & $5.869 \times 10^{-4}$  & $5.54  \times 10^{-3}$  &             & $0.020$ & $7.009 \times 10^{-4}$ & $3.097$ \\ \hline
           & $0.10$   & $9.000 \times 10^{-4}$  & $5.54  \times 10^{-3}$  &             & $0.10$   & $9.000 \times 10^{-4}$ & $3.097$ \\ \hline
           & $0.20$   & $1.299 \times 10^{-3}$  & $5.54  \times 10^{-3}$  &             & $0.20$   & $1.111 \times 10^{-3}$ & $3.097$ \\ \hline 
           & $1.0$     & $4.972 \times 10^{-3}$  & $5.54  \times 10^{-3}$  &             & $1.0$     & $2.824 \times 10^{-3}$ & $3.097$ \\ \hline \hline
$5$     & ---          & ---                                   & $1.38 \times 10^{-1}$  & $40$      & --- & --- & 5.61 \\ \hline                             
           & $0.020$ & $6.584 \times 10^{-4}$  & $0.328$ &                  & $0.020$ & $7.18 \times 10^{-4}$ & $8.08$ \\ \hline
           & $0.10$   & $9.000 \times 10^{-4}$  & $0.328$ &                  & $0.10$   & $9.000 \times 10^{-4}$ & $8.08$ \\ \hline
           & $0.20$   & $1.177 \times 10^{-3}$  & $0.328$ &                  & $0.20$   & $1.084 \times 10^{-3}$ & $8.08$ \\ \hline
           & $1.0$     & $3.632 \times 10^{-3}$  & $0.328$ &                  & $1.0$     & $2.513 \times 10^{-3}$ & $8.08$ \\ \hline \hline
\end{tabular}
\caption{Parameters for the same value of the cross sections for  $\nu_\mu \to \mu_\mu \mu \Am$. The units of $E_\nu$ and $m_{Z'}$ are GeV, 
and that of $\sigma$ is $10^{-41}$ cm$^2$, respectively.}
\label{tab:parameter-set-const-cs-mu}
\end{table}
%%%%%%%%%%%%%%%%%%%%%%%%%%%%%%%%
For this purpose, we analyze the distributions in the energies, opening angle and invariant mass 
of the final state charged leptons in the next section. 
The analyses are performed on the parameter sets shown in Table \ref{tab:parameter-set-const-cs-mu} for  
the muonic trident. 
The first row for each $E_\nu$ in Table \ref{tab:parameter-set-const-cs-mu} is the trident cross section in the SM. 
We chose $(\mZp, g') = (0.1$ GeV$, 9 \times 10^{-4})$ as a reference parameter, which can explain $(g-2)_\mu$ 
within $2\sigma$. 
Other parameter sets are chosen so that the cross sections have the same values with that of the reference set for each $E_\nu$. 
Note that some parameter sets are outside the $2\sigma$ region of $(g-2)_\mu$ or in the gray region. 
However, we include those parameter sets to see the behavior of the distributions for comparison.

%%%%%%%%%%%%%%%%%%%%%%%%%%%%%%%%
\begin{table}[t]
\begin{tabular}{|c|c|c|c||c|c|c|c|} \hline
~~$E_\nu$~~ & ~~~ $m_{Z'}$ ~~~ & ~~~~~~ $g'$ ~~~~~~ & ~~~~~~~~ $\sigma$ ~~~~~~~~ &
~~$E_\nu$~~ & ~~~ $m_{Z'}$ ~~~ & ~~~~~~ $g'$ ~~~~~~ & ~~~~~~~~ $\sigma$ ~~~~~~~~ \\ \hline \hline
$10$   & ---           & ---                              & $1.94 \times 10^{-8}$  &  $40$ & ---           & ---                              & $1.87$ \\ \hline
           & $0.020$ & $9.53\times 10^{-4}$ & $1.98 \times 10^{-8}$  &           & $0.020$ & $9.09\times 10^{-4}$ & $1.87$ \\ \hline
           & $0.10$   & $9.00 \times 10^{-4}$  & $1.96 \times 10^{-8}$ &          & $0.10$   & $9.00 \times 10^{-4}$ & $1.87$ \\ \hline
           & $0.20$   & $9.29\times 10^{-4}$ & $1.95 \times 10^{-8}$   &          & $0.20$   & $9.13\times 10^{-4}$ & $1.87$ \\ \hline
           & $1.0$     & $1.47\times 10^{-3}$ & $1.94 \times 10^{-8}$  & & $1.0$  & $1.20\times 10^{-3}$ & $1.87$ \\ 
           \hline \hline
\end{tabular}
\caption{Parameters for the same value of the cross section for $\nu_\mu \to \nu_\mu \tau \At $. The units of $E_\nu$ and $m_{Z'}$ are GeV, 
and that of $\sigma$ is $10^{-47}$ cm$^2$, respectively.}
\label{tab:parameter-set-const-cs-tau}
\end{table}
%%%%%%%%%%%%%%%%%%%%%%%%%%%%%%%%
We also show the cross sections of tauonic NTP, $\nu_\mu \to \nu_\mu \tau \At$, in Table \ref{tab:parameter-set-const-cs-tau} 
for $E_\nu = 10$ and $40$ GeV. One finds that the cross sections for the reference point are almost the same as that in the SM. 
This suggests that the new physics contributions are very small. For tauonic trident to occur, the momentum transfer 
$|q_{Z'}^2|$ will be of order $m_\tau^2$ and hence the new physics contribution is much suppressed as shown in
 Eq.~\eqref{eq:ntp-cond}.
In fact, we have performed the same analyses for the tauonic NTP as for muonic one in the next section, 
and found that the distributions show tiny difference among the NP parameter sets in Table \ref{tab:parameter-set-const-cs-tau}. 
Therefore, we show our numerical results only for muonic NTP in the next section.

%%%%%%%%%%%%%%%%%%%%%%%%%%%%%%%%%%%%%%%%%%%%%%%%%%%%%%%%%%
\section{Numerical Results}
%%%%%%%%%%%%%%%%%%%%%%%%%%%%%%%%%%%%%%%%%%%%%%%%%%%%%%%%%%
We show the distributions of the energies $E_\mu$ and $E_{\Am}$, invariant mass $m^2_{\mu\Am}$ and opening angle $\theta_{\mu\Am}$ of 
muon and anti-muon in the SM and our model for the parameters given in Table \ref{tab:parameter-set-const-cs-mu}.
To obtain the total cross section of the NTP processes, we have to perform the phase space integral with a seven-dimension.
For such high-dimensional integrals, the Monte Carlo integration is known to be useful due to its quick convergence compared to quadratures by parts.

To investigate the NP effect in the charged lepton distributions, we calculate the differential cross section with respect to some observables.
In general, it is complicated to select an arbitrary observable as one of integral variables.
However, when we use the Monte Carlo integration, we do not need to make the complicated variable transformation to obtain the differential cross section with respect to the favored observable.

Let $f(\bm{y})$ be a function of variables $\bm{y}$, which satisfies $\sigma=\int d\bm{y}f(\bm{y})$.
Here, treating $\bm{y}$ as integral variables, we consider to perform the Monte Carlo integration.
To obtain $d\sigma/dx$, we prepare discretized bins of a variable $x$, which are labeled by $a$ and have an interval $\Delta x_a$.
In this integration, we sample the variables $\bm{y}$ from the uniform probability distribution $N$ times.
At $i$-th step of the sampling, $x_i$ is calculated as well as $f(\bm{y}_i)$ for generated $\bm{y}_i$.
Then, one can approximate the distribution in the variable $x$ by
\begin{align}
\frac{d \sigma}{d x}(x_a) \simeq \frac{D}{N\Delta x_a}\sum_{i=1}^{N}f(\bm{y}_i)\theta\left(x_i-x_a+\frac{\Delta x_a}{2}\right)\theta\left(x_a+\frac{\Delta x_a}{2}-x_i\right),
\label{eq:diff-mc}
\end{align}
where $D$ is the total width of the $x$ bins and $N$ is the number of samples.
The function $\theta(z)$ is a step function, which is a unity for $z\ge 0$ and zero for $z<0$.
The total cross section can be obtained by summing Eq.~\eqref{eq:diff-mc} over $x$ as
\begin{align}
\sigma \simeq \sum_{a} \Delta x_a \frac{d\sigma}{d x}(x_a).
\end{align}

%%%%%%%%%%%%%%%%%%%%%%%%%%%%%%%%%%%%%%%%%%%%%%%%%%%%%%%%%%
\subsection{Energy Distributions}
%%%%%%%%%%%%%%%%%%%%%%%%%%%%%%%%%%%%%%%%%%%%%%%%%%%%%%%%%%

%%%%%%%%%%%%%%%%%%%%%%%%%%%%%%%%%%%%%%%
\begin{figure}[t]
\begin{tabular}{ccc}
	\includegraphics[width=0.5\linewidth]{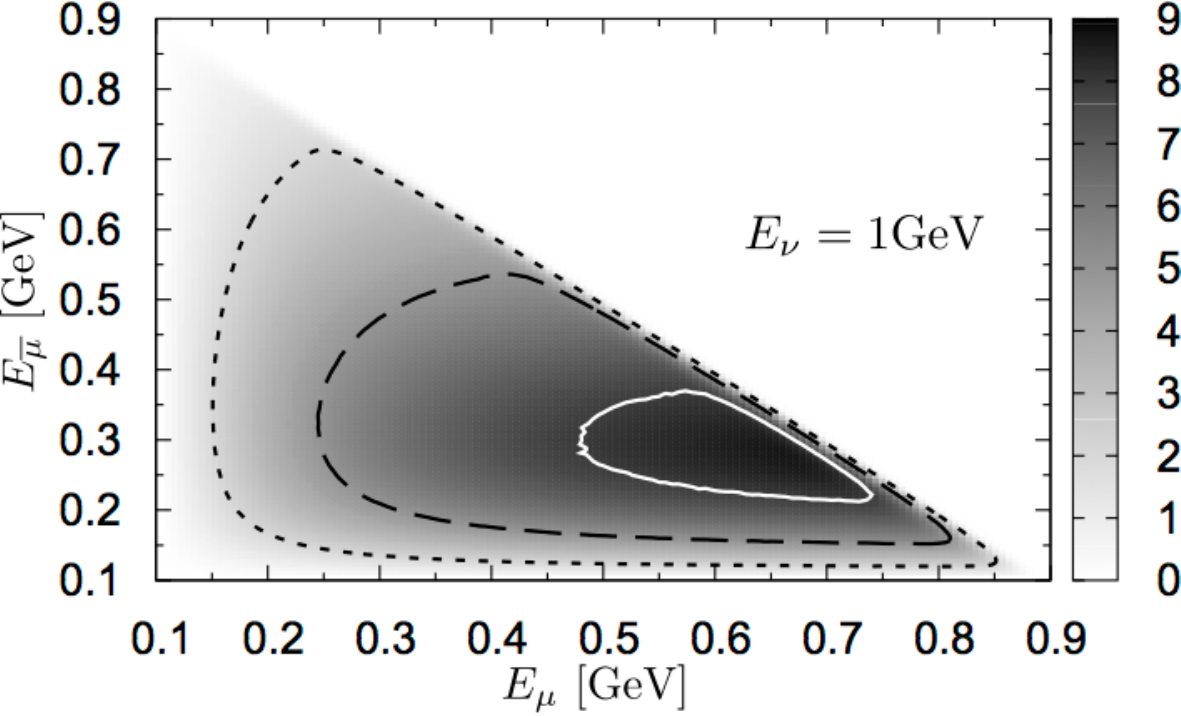} & ~~& 
	\includegraphics[width=0.5\linewidth]{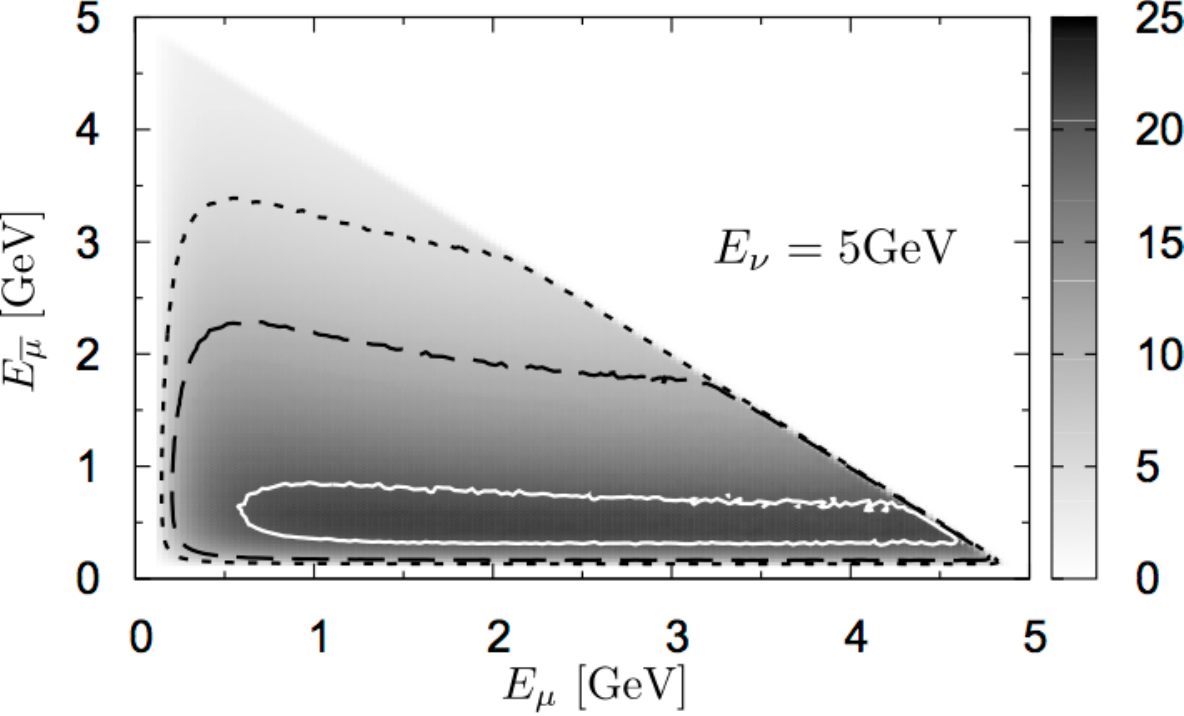} \\
	\includegraphics[width=0.5\linewidth]{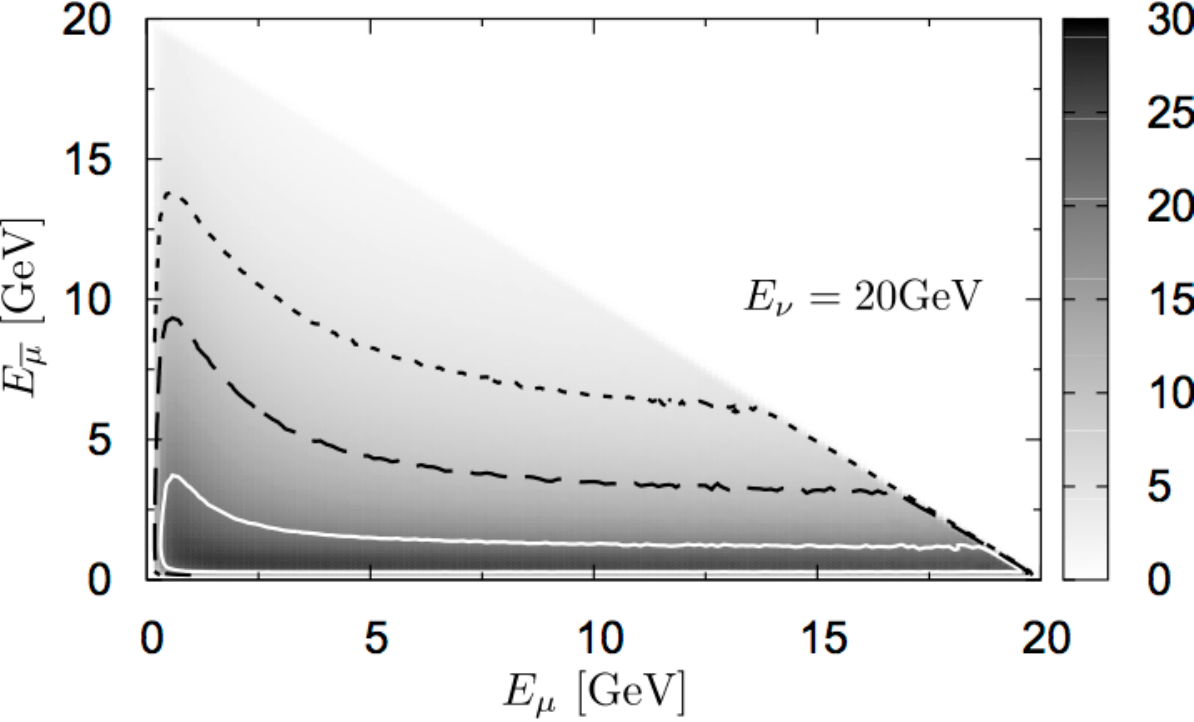} & ~~&
	\includegraphics[width=0.5\linewidth]{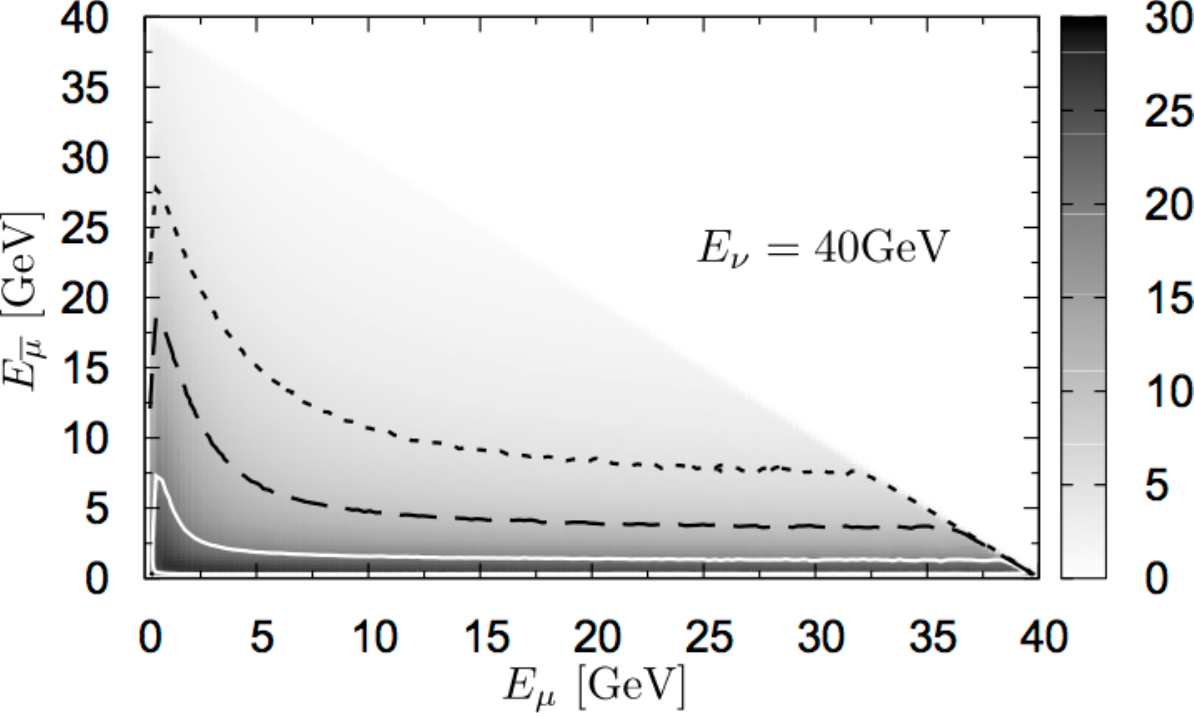} 
\end{tabular}
\caption{Contour plots of the distribution in $E_\mu$-$E_{\bar{\mu}}$ plane for $\nu_\mu \to \nu_\mu \mu \Am$ in the SM. 
The color bar and contours are in unit of $10^{-44}$cm$^2$/GeV$^2$. 
Solid (white), dashed (black) and dotted (black) curves are the contours of $8,~5$ and $2$ for $E_\nu = 1$ GeV, 
and $20,~10$ and $5$ for $E_\nu = 5,~20,~40$ GeV, respectively.
}
\label{fig:energy-dist-SM}
\end{figure}
%%%%%%%%%%%%%%%%%%%%%%%%%%%%%%%%%%%%%%%
%
Firstly, we show the SM distributions of $E_\mu$ and $E_{\Am}$ for the process of $\nu_\mu \to \nu_\mu \mu \Am$
in Fig.~\ref{fig:energy-dist-SM}.  The energy of the incoming neutrino is taken to be $E_\nu = 1,~5,~20$ and $40$ GeV, 
respectively.
The color bar on the right indicates the value of the double differential cross section 
in unit of $10^{-44}$ cm$^2$/GeV$^2$.
Solid (white), dashed (black) and dotted (black) curves are the contours of $8,~5$ and $2$ in $10^{-44}$ cm$^2$/GeV$^2$
for $E_\nu = 1$ GeV, while $20,~10$ and $5$ in $10^{-44}$ cm$^2$/GeV$^2$ for $E_\nu = 5,~20,~40$ GeV, respectively.

In each panel, one can see that the distribution has a peak near the kinematical edge for $E_\nu = 1$ GeV.  
As $E_\nu$ becomes higher, the peak moves to lower $E_\mu$ region. 
For $E_\nu > 5$ GeV, $E_\mu$ is uniformly distributed rather than $E_{\Am}$ is.
We can understand this asymmetry of the distribution in $E_\mu$-$E_{\Am}$ plane as follows: As we explained in Eq.~\eqref{eq:ML-MR-exchange}, the terms $M_L$ and $M_R$ in the lepton tensor are exchanged
under $p \leftrightarrow \overline{p}$. 
Thus, the double differential cross section differs under the exchange of $E_\mu \leftrightarrow E_{\Am}$ if the coupling constants $g_L$ and $g_R$ are different as in the SM. It should be noticed 
that the distribution becomes symmetric in $E_\mu$-$E_{\Am}$ plane for the case of $g_L = g_R$, such that 
the $\lmlt$ contributions dominate over the SM couplings.
%

%%%%%%%%%%%%%%%%%%%%%%%%%%%%%%%%%%%%%%%
\begin{figure}[t]
\begin{tabular}{ccc}
	\includegraphics[width=0.49\linewidth]{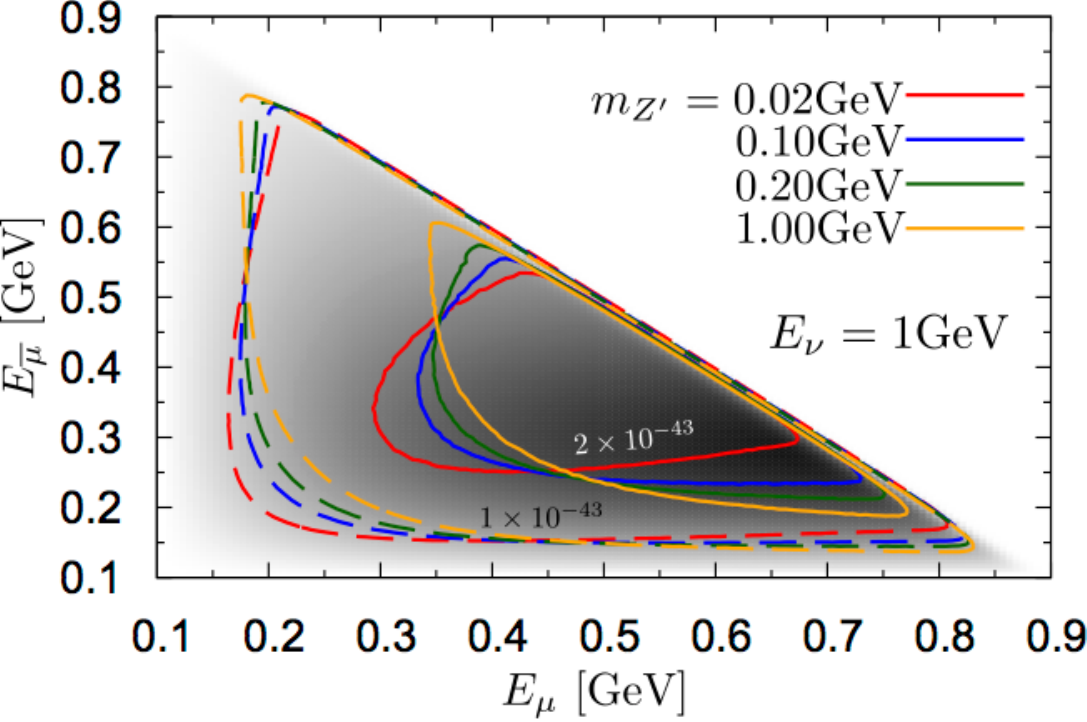} & ~~& 
	\includegraphics[width=0.49\linewidth]{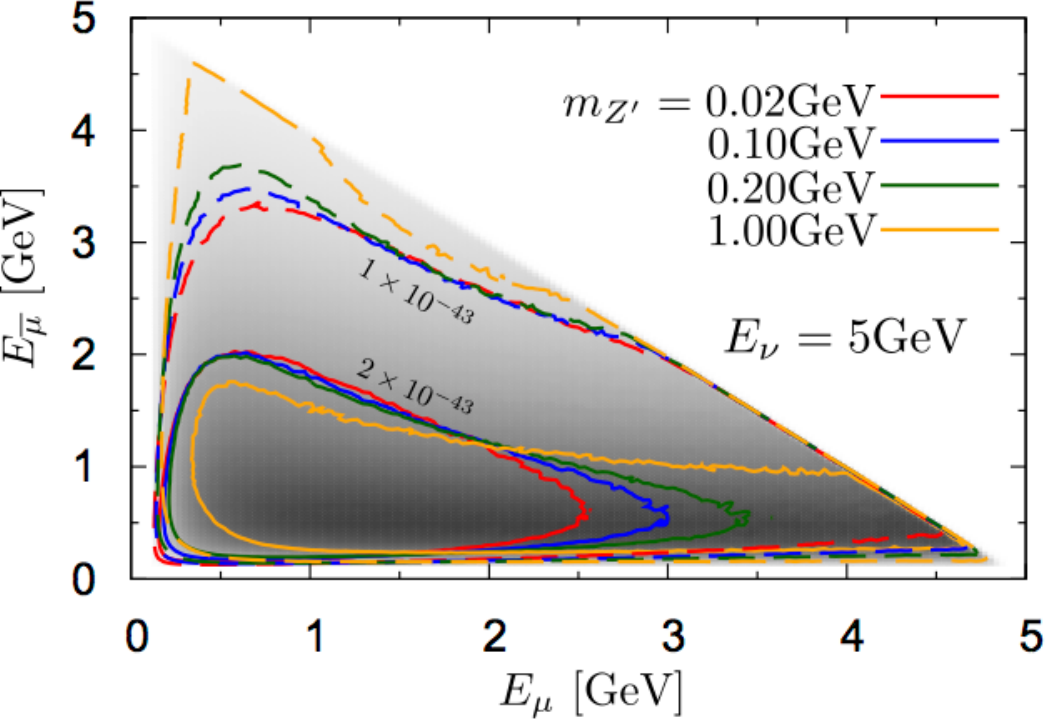} \\
	\includegraphics[width=0.49\linewidth]{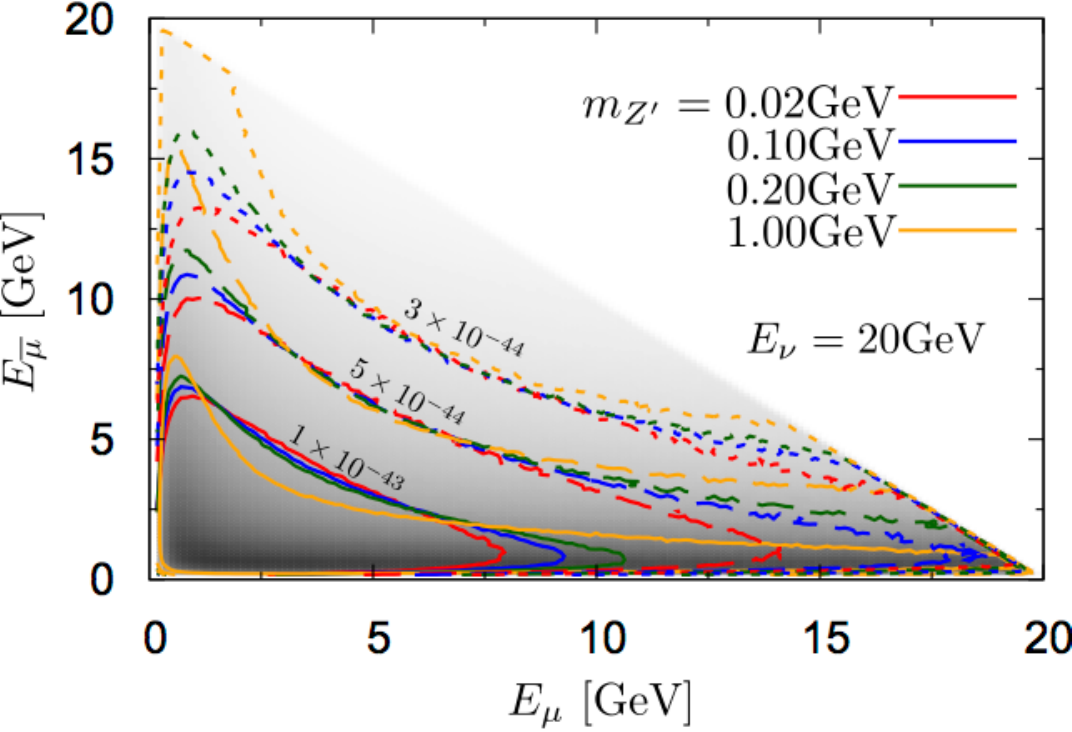} & ~~&
	\includegraphics[width=0.49\linewidth]{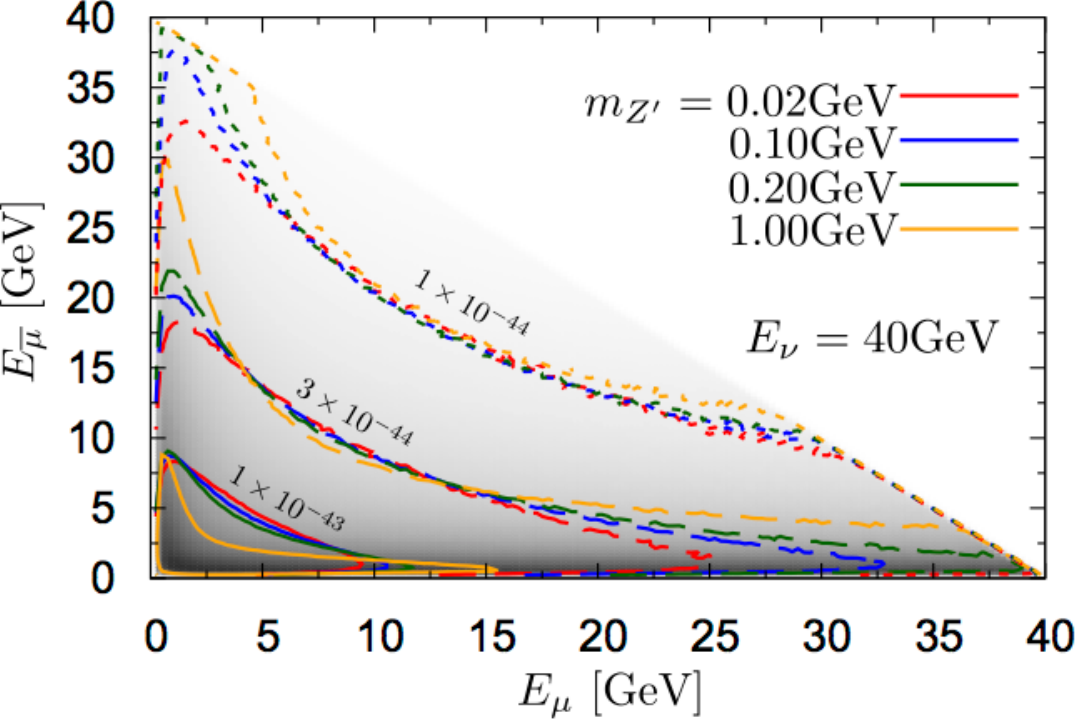} 
\end{tabular}
\caption{Contour plots of the deviation from the SM in the $E_\mu$-$E_{\bar{\mu}}$ distribution of 
$\nu_\mu \to \nu_\mu \mu \Am$. 
Red, blue, green and orange curves correspond to the parameter sets in Table \ref{tab:parameter-set-const-cs-mu}. 
Solid, dashed and dotted curves are the contours of $R(E_\mu, E_{\Am})$ with values indicated near each curve. 
Background gray color is the SM distribution shown in Fig.~\ref{fig:energy-dist-SM}.}
\label{fig:energy-dist-NP-SM}
\end{figure}
%%%%%%%%%%%%%%%%%%%%%%%%%%%%%%%%%%%%%%%
%
To see the parameter dependence of the NP model, we show the deviation of the differential cross section 
in our model from the SM, which is defined by
\begin{align}
R(E_\mu, E_{\Am}) \equiv 
 \frac{d^2 \sigma}{d E_\mu d E_{\Am}}  -  \frac{d^2 \sigma_\mathrm{SM}}{d E_\mu d E_{\Am}}.
\end{align}
In Fig.~\ref{fig:energy-dist-NP-SM},  solid, dashed and dotted curves are the contours of $R(E_\mu, E_{\Am})$ while  
red, blue, green and orange colors represent the parameter sets in Table \ref{tab:parameter-set-const-cs-mu}. 
In each panel, the values of $R(E_\mu, E_{\Am})$ are indicated near each curve, and only the $Z'$ mass is 
shown to specify the parameter sets.  The gray background represents the SM distribution shown in 
Fig.~\ref{fig:energy-dist-SM}.

From Fig.~\ref{fig:energy-dist-NP-SM}, the parameter dependence can be seen clearly in the contours of $R(E_\mu, E_{\Am})$ 
in high $E_{\mu}$ or $E_{\Am}$ region . 
The contours extend to larger values of $E_\mu$ or $E_{\Am}$ as the $Z'$ mass is heavier. 
It is also seen that $E_\mu$ is uniformly distributed rather than $E_{\Am}$ is, due to the interference between the 
NP and SM contributions. We note that $R(E_\mu, E_{\Am})$ is positive for all ($E_\mu,E_{\Am}$) in $\nu_\mu \to \nu_\mu \mu \Am$ in our calculation.
The $\lmlt$ contribution to enhances $\nu_\mu \to \nu_\mu \mu \Am$ because both $g_L$ and $g_R$ are effectively enlarged by the propagator of the $Z'$ boson as seen in Eq.~\eqref{eq:coupling}.

%%%%%%%%%%%%%%%%%%%%%%%%%%%%%%%%%%%%%%%
\begin{figure}[t]
\includegraphics[width=\linewidth]{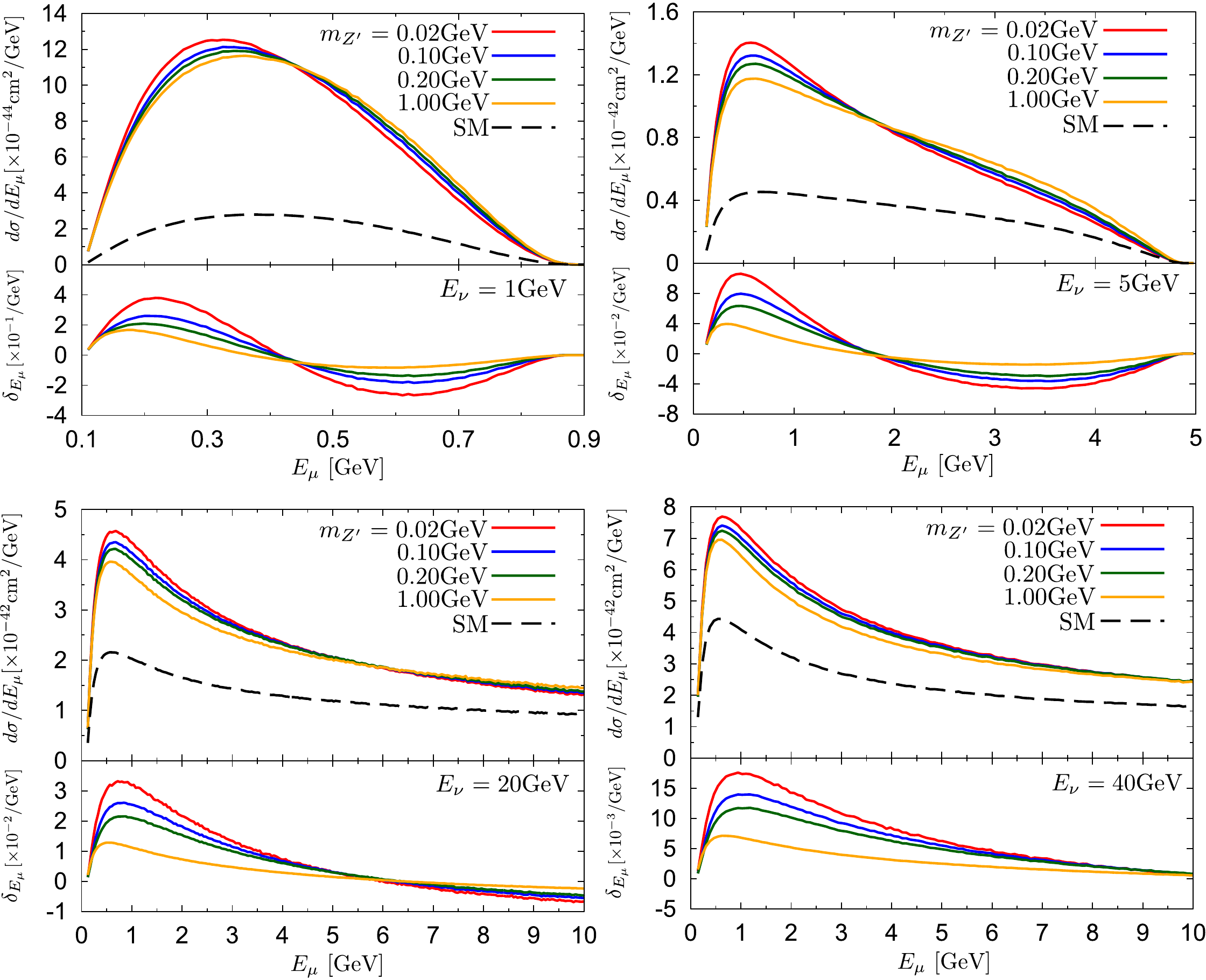}
\caption{The $E_\mu$ distribution (upper panel) and shape of the distribution in $\lmlt$ model. 
Color of the curves are the same as Fig.~\ref{fig:energy-dist-NP-SM}. Black dashed curve represents the 
SM distribution.}
\label{fig:NP-Emu-dist}
\end{figure}
\begin{figure}[t]
\includegraphics[width=\linewidth]{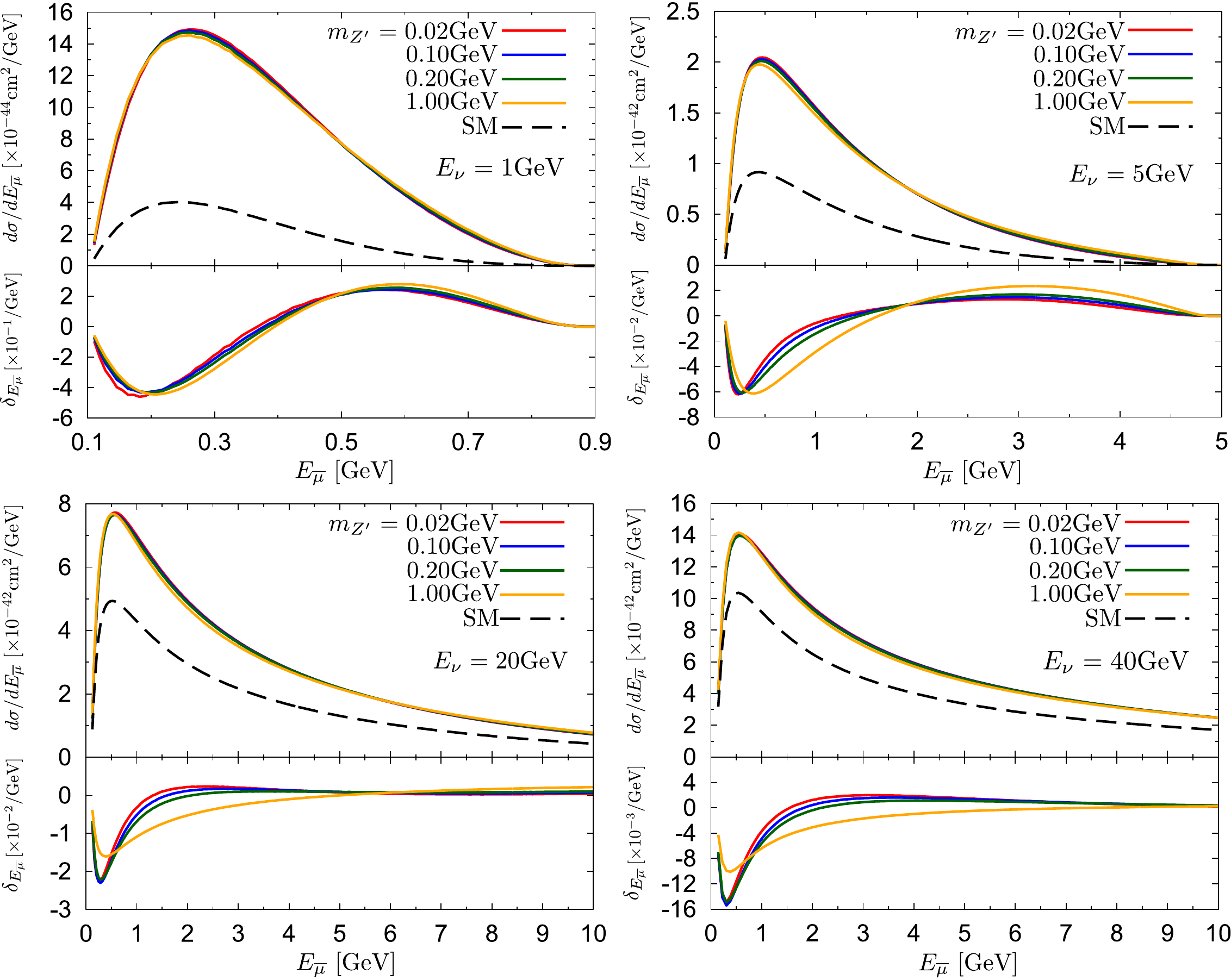}
\caption{$E_{\Am}$ distributions in $\lmlt$ model.}
\label{fig:NP-Emubar-dist}
\end{figure}

%%%%%%%%%%%%%%%%%%%%%%%%%%%%%%%%%%%%%%%
Figure~\ref{fig:NP-Emu-dist} and \ref{fig:NP-Emubar-dist} show that the distributions (upper panel) and difference of 
shape of the distribution from the SM (lower panel) in $E_\mu$ and $E_{\Am}$, respectively. 
Colors of the solid curves are the same in Fig.~\ref{fig:energy-dist-NP-SM} and black dashed curve is the SM distribution.
In the upper panels of Fig.~\ref{fig:NP-Emu-dist}, the parameter dependence of the distributions can be seen 
in two regions, around the peaks in lower $E_\mu$ and at tails in higher $E_\mu$. The parameter dependence is 
clearer around the peaks than at the tail. In each panels, one can see that the peaks become higher as the $Z'$ mass is lighter. 
It is also seen that $E_\mu$ corresponding to the peak slightly differs among the parameter sets in each energy.
On the other hand, in the upper panels of Fig.~\ref{fig:NP-Emubar-dist}, the $E_{\Am}$ distribution is less dependent on the 
parameters compared with the $E_\mu$ distributions.
These results imply that the $E_\mu$ distribution is more useful to determine the new physics parameters.

%%%%%%%%%%%%%%%%%%%%%%%%%%%%%%%%%%%%%%%
%%%%%%%%%%%%%%%%%%%%%%%%%%%%%%%%%%%%%%%
%
To see how the NP contributions modify the shape of the distributions, 
we define
\begin{align}
\delta_{X} \equiv \frac{1}{\sigma} \frac{d\sigma}{d X} 
- \frac{1}{\sigma_{\mathrm{SM}}} \frac{d\sigma_{\mathrm{SM}}}{d X}. \label{eq:diff-Emu}
\end{align}
for an arbitrary kenematical variable $X$.
In the lower panels of Figs.~\ref{fig:NP-Emu-dist} and \ref{fig:NP-Emubar-dist}, we plotted $\delta_{E_\mu}$ and $\delta_{E_{\Am}}$, respectively.
These are the difference of the normalized 
distributions, and become zero when the shape of the distributions are the same between our model and the SM, 
even if overall magnitudes are different.  
One can see in Fig.~\ref{fig:NP-Emu-dist} that $\delta_{E_\mu}$ is positive in lower $E_\mu$ and negative in higher $E_\mu$ 
for all parameter sets. 
On the other hand, in Fig.~\ref{fig:NP-Emubar-dist}, $\delta_{E_{\Am}}$ shows the opposite behavior. 
Thus, the distributions are shifted to the lower $E_\mu$ and higher $E_{\Am}$ by the NP contribution.
The shape of the distribution also depends on the NP parameter set. In the $E_\mu$ distribution, $|\delta_{E_\mu}|$ 
is larger for the lighter $Z'$ mass. Such information can be used to determine the parameters.

%%%%%%%%%%%%%%%%%%%%%%%%%%%%%%%%%%%%%%%%%%%%%%%%%%%%%%%%%%
\subsection{Invariant Mass Distributions}
%%%%%%%%%%%%%%%%%%%%%%%%%%%%%%%%%%%%%%%%%%%%%%%%%%%%%%%%%%

%%%%%%%%%%%%%%%%%%%%%%%%%%%%%%%%%%%%%%%
\begin{figure}[t]
\includegraphics[width=\linewidth]{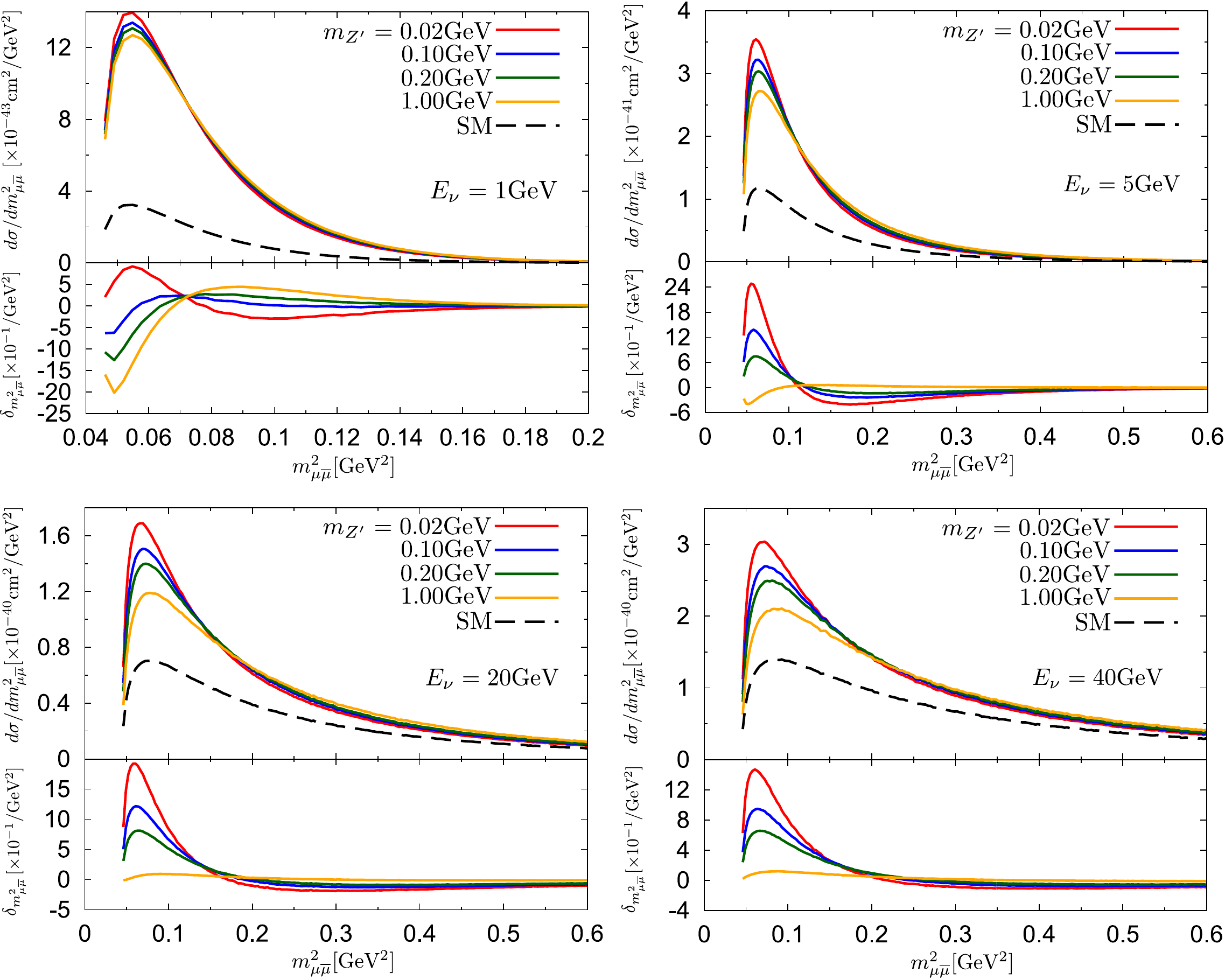} 
\caption{$m_{\mu\overline{\mu}}^2$ distributions in $L_\mu-L_\tau$ model.}
\label{fig:invariant-mass-dist-NP}
\end{figure}
%%%%%%%%%%%%%%%%%%%%%%%%%%%%%%%%%%%%%%%
%
The invariant mass of the outgoing muon and anti-muon is defined by
\begin{align}
m^2_{\mu \Am} \equiv (p + \overline{p})^2.
\end{align}
In Fig.~\ref{fig:invariant-mass-dist-NP}, we show the invariant mass distribution (upper panel) and shape difference of this distribution from the SM
(lower panel). 

We can see from each panel that the distribution clearly depends on the parameters in 
lower value of $m^2_{\mu\Am}$ region. The peaks of the distributions become sharper as the $Z'$ mass is lighter and 
the dependence becomes more significant as $E_\nu$ is higher. It is also seen that the $m^2_{\mu\Am}$ 
corresponding to the peak changes for the parameter sets.
From the lower panels, one can also see that $\delta_{m^2_{\mu\Am}}$, defined by Eq.~\ref{eq:diff-Emu}, can take positive and negative values  in lower value of $m^2_{\mu\Am}$ depending on the parameters, which 
shows different behaviors from the energy distributions.

%%%%%%%%%%%%%%%%%%%%%%%%%%%%%%%%%%%%%%%%%%%%%%%%%%%%%%%%%%
\subsection{Opening Angle Distributions}
%%%%%%%%%%%%%%%%%%%%%%%%%%%%%%%%%%%%%%%%%%%%%%%%%%%%%%%%%%

%%%%%%%%%%%%%%%%%%%%%%%%%%%%%%%%%%%%%%%
\begin{figure}[t]
\includegraphics[width=\linewidth]{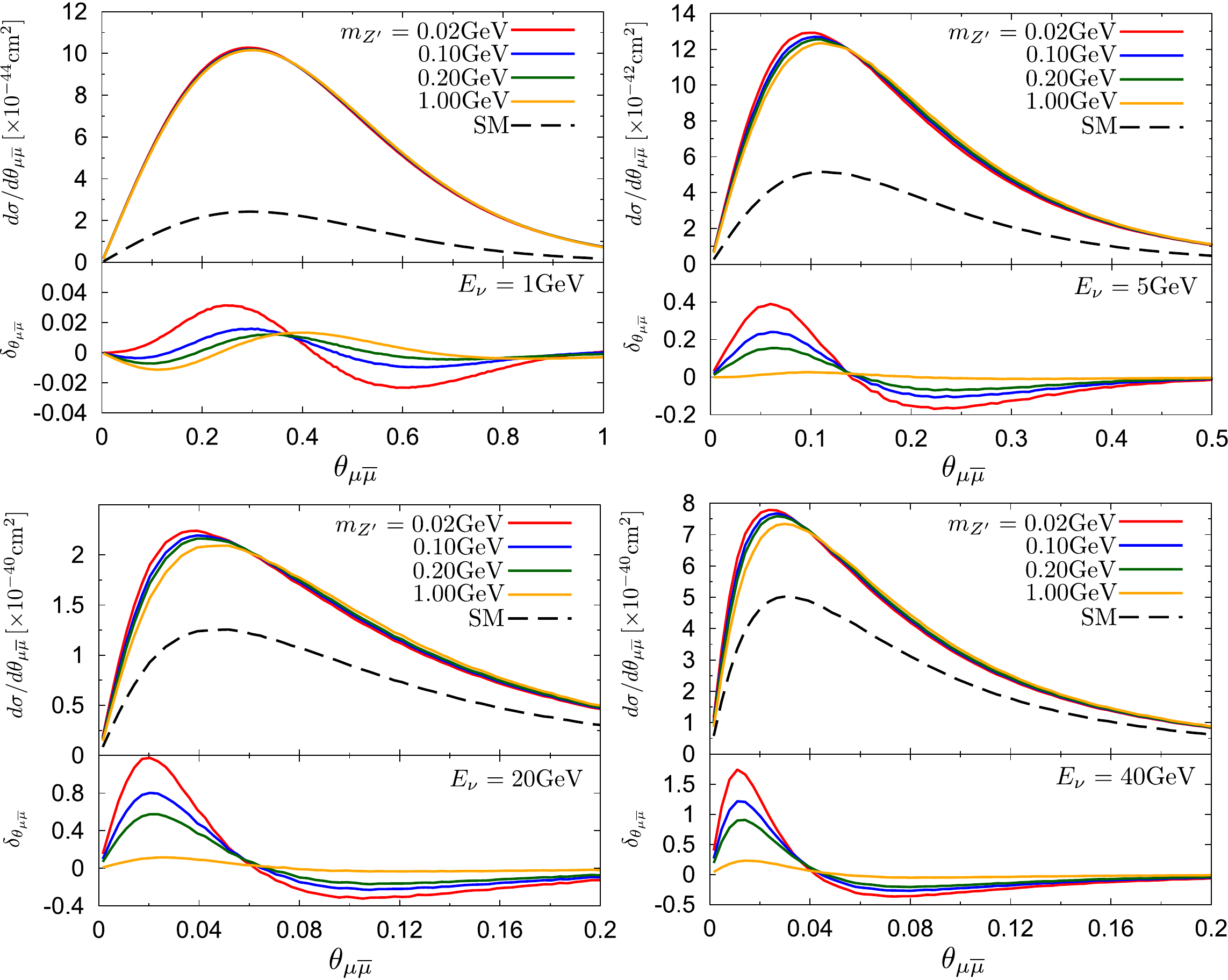}
\caption{Muon opening angle distributions of the cross section in the $L_\mu-L_\tau$ model.}
\label{fig:angular-dist-NP}
\end{figure}
%%%%%%%%%%%%%%%%%%%%%%%%%%%%%%%%%%%%%%%
%
The opening angle of the outgoing muon and anti-muon can be defined by
\begin{align}
\cos\theta_{\mu\Am} \equiv \frac{\bm{p} \cdot \overline{\bm{p}}}{|\bm{p}| |\overline{\bm{p}}|},
\end{align}
where $\bm{p}$ and $\overline{\bm{p}}$ are the momentum of muon and anti-muon, respectively.
Figure~\ref{fig:angular-dist-NP} shows the distributions of $\theta_{\mu\Am}$ (upper panel) and the shape (lower panel). 
From figure, one finds that the opening angle distributions shows the parameter dependence around the peaks 
for higher $E_\nu$. It is also seen that the angle for the peaks becomes smaller as $E_\nu$ becomes higher.
Similar behavior can be seen in the shape of the distributions.

In this analysis, we have considered only the coherent NTP processes assuming argon as target material.
Since the kinematical distributions depend on the form factor, the nuclear dependence is worth investigating for finding the best target material to identify the NP parameters.

We also understand that the diffractive NTP processes can be relevant for the available neutrino energy in experiments.
We should check whether the diffractive contribution makes positive or negative effects for the measurement of NP parameters.
We will study these topics in our future works.

%%%%%%%%%%%%%%%%%%%%%%%%%%%%%%%%%%%%%%%%%%%%%%%%%%%%%%%%%%
\section{Summary}
%%%%%%%%%%%%%%%%%%%%%%%%%%%%%%%%%%%%%%%%%%%%%%%%%%%%%%%%%%
We have considered the minimal gauged $\lmlt$ model, and studied the dependences of the distributions of 
the neutrino trident production process on the new physics parameters, $\mZp$ and $g'$.
We analyzed the distributions of energies, opening angle and invariant mass of muons 
and their shapes in $\nu_\mu \to \nu_\mu \mu \Am$. 

We have found that the distributions can be different among the NP parameter sets 
for which the total cross sections are the same. 
In $E_\mu$-$E_{\Am}$ distributions, the differences can be seen in larger $E_\mu$ or $E_{\Am}$ region. 
We also found that the parameter dependences in the $E_\mu$ and $m_{\mu\Am}^2$ distributions are rather clear compared 
with those in the $E_{\Am}$ and $\theta_{\mu\Am}$ distributions. 
Therefore the $E_\mu$ and $m_{\mu\Am}^2$ distributions will be useful to determine the NP parameters.
The shapes of the distributions were also presented, which shows the parameter dependence. 

The determination of the new physics parameters by combining the information of the total cross section and 
distributions will be next step of our work. Such a study will need more detailed information  
including resolution and efficiencies in experiments. We leave such analyses for next work.

\begin{acknowledgements}
This work is supported by JSPS KAKENHI Grant Number JP18K03651 and MEXT KAKENHI Grant Number 
JP18H05543 (T.~S.), the Sasakawa Scientific Research Grant from the Japan 
 Science Society (Y.~U.), and JSPS KAKENHI Grant Number JP18H01210 (T.~S. and Y.~U.).
\end{acknowledgements}

%%%%%%%%%%%%%%%%%%%%%%%%%%

\appendix

%%%%%%%%%%%%%%%%%%%%%%%%%%%%%%%%%%%%%%%%%%%%%%%%%%%%%%%%%%
\section{Lepton Tensor} \label{apdx:clep-tensor}
%%%%%%%%%%%%%%%%%%%%%%%%%%%%%%%%%%%%%%%%%%%%%%%%%%%%%%%%%%
In this appendix, we present the analytic formula of the amplitude squared for $\nu_\mu\to\nu_\mu\ell\overline{\ell}$. 
The lepton and nucleus parts are written as a neutrino tensor $j^{\alpha\beta}$, charged lepton tensor $L_{\alpha\beta}^{\mu\nu}$, and nucleus tensor $J_{\mu\nu}$.
Then, the amplitude squared is given by $j^{\alpha \beta} L^{\mu \nu}_{\alpha\beta}J_{\mu \nu}$ up to overall factors. 
Using the expression of the charged lepton tensor, \eqref{seq:charged-part}, it can be 
classified by the chirality of outgoing charged leptons as
\begin{align}
j^{\alpha\beta}L_{\alpha\beta}^{\mu\nu}J_{\mu\nu}=&\left|g_L\right|^2M_L+\left|g_R\right|^2M_R-g_Lg_R^*M_{LR}-g_L^*g_RM_{RL},
\end{align}
where
\begin{align}
M_L=& 4J_{\mu\nu}j^{\alpha\beta}\mathrm{Tr}\left[\Slash{p}W_\alpha^\mu\Slash{\overline{p}}V_\beta^\nu P_L\right], \\
M_R=& 4J_{\mu\nu}j^{\alpha\beta}\mathrm{Tr}\left[\Slash{p}W_\alpha^\mu\Slash{\overline{p}}V_\beta^\nu P_R\right], \\
M_{LR}=& 4m_\ell^2J_{\mu\nu}j^{\alpha\beta}\mathrm{Tr}\left[W_\alpha^\mu V_\beta^\nu P_R\right], \\
M_{RL}=& 4m_\ell^2J_{\mu\nu}j^{\alpha\beta}\mathrm{Tr}\left[W_\alpha^\mu V_\beta^\nu P_L\right].
\end{align}
Here, $W^\mu_\alpha$ and $V^\nu_\beta$ are the propagators of leptons given by
\begin{align}
W_\alpha^\mu=&\frac{2p^\mu+\gamma^\mu\Slash{q}}{q^2+2p\cdot q}\gamma_\alpha-\gamma_\alpha\frac{2\overline{p}^\mu+\Slash{q}\gamma^\mu}{q^2+2\overline{p}\cdot q}.
\end{align}
and
\begin{align}
V_\beta^\nu=&\gamma_0W_\beta^{\dagger\nu}\gamma_0=\gamma_\beta\frac{2p^\nu+\Slash{q}\gamma^\nu}{q^2+2p\cdot q}-\frac{2\overline{p}^\nu+\gamma^\nu\Slash{q}}{q^2+2\overline{p}\cdot q}\gamma_\beta.
\end{align}
Due to the parity conservation of the electomagnetic interaction, $J_{\mu\nu}$ must be symmetric with respect to $\mu$ and $\nu$, regardless of the detail 
of the nucleus.
Thus, it is enough to calculate the symmetric part of  $j^{\alpha\beta}L_{\alpha\beta}^{\mu\nu}$ under $\mu \leftrightarrow \nu$.
The concrete form of the nucleus tensor $J_{\mu\nu}$ is determined by the target nucleus, as given in Eq.~\eqref{eq:Jmunu} with \eqref{eq:nuc-density}.

By the straightforward calculation, we obtain the explicit formula for $M_L$ in terms of lepton momenta as follows:
\begin{align}
M_L=&(-256J_{\mu\nu})\left\{\frac{M_{L1}^{\mu\nu}}{\left(q^2+2p\cdot q\right)^2}+\frac{M_{L2}^{\mu\nu}}{\left(q^2+2\overline{p}\cdot q\right)^2}+\frac{M_{L3}^{\mu\nu}}{\left(q^2+2p\cdot q\right)\left(q^2+2\overline{p}\cdot q\right)}\right\}, \\
M_{L1}^{\mu\nu}=& k\cdot\overline{p}\left[g^{\mu\nu}\left(2q\cdot k'q\cdot p-q^2k'\cdot p\right)+\left(p^\mu k'^\nu+p^\nu k'^\mu\right)\left(q^2+2p\cdot q\right) \right. \nonumber\\
&\hspace{1cm}\left.-2\left(p^\mu q^\nu+p^\nu q^\mu+2p^\mu p^\nu\right)\left\{k'\cdot\left(p+q\right)\right\}\right], \\
M_{L2}^{\mu\nu}=&\left(\text{$\{p,k\}\leftrightarrow\{\overline{p},k'\}$ exchange of $M_{L1}^{\mu\nu}$}\right), \\
M_{L3}^{\mu\nu}=& 2g^{\mu\nu}\left\{p\cdot q\left(k'\cdot\overline{p}q\cdot k-q\cdot\overline{p}k\cdot k'\right)\right. \nonumber\\
&\hspace{1cm}\left.+q\cdot k'\left(q\cdot\overline{p}k\cdot p-p\cdot\overline{p}q\cdot k\right)+q^2\left(p\cdot\overline{p}k\cdot k'-p\cdot k\overline{p}\cdot k'\right)\right\} \nonumber\\
&+\left(p^\mu\overline{p}^\nu+p^\nu\overline{p}^\mu\right)\left(4k\cdot\overline{p}k'\cdot p+2k\cdot qk'\cdot q-q^2k\cdot k'+2k\cdot\overline{p}q\cdot k'+2k'\cdot pq\cdot k\right) \nonumber\\
&+\left(p^\mu q^\nu+p^\nu q^\mu\right)\left(k\cdot\overline{p}q\cdot k'+q\cdot\overline{p}k\cdot k'-q\cdot kk'\cdot\overline{p}+2k'\cdot pk\cdot\overline{p}\right) \nonumber\\
&+\left(\overline{p}^\mu q^\nu+\overline{p}^\nu q^\mu\right)\left(k'\cdot pq\cdot k+q\cdot pk\cdot k'-q\cdot k'k\cdot p+2k\cdot\overline{p}k'\cdot p\right) \nonumber\\
&-\left(p^\mu k^\nu+p^\nu k^\mu\right)\left(2k'\cdot pq\cdot\overline{p}+2q\cdot\overline{p}q\cdot k'-q^2k'\cdot\overline{p}\right) \nonumber\\
&-\left(\overline{p}^\mu k'^\nu+\overline{p}^\nu k'^\mu\right)\left(2k\cdot\overline{p}q\cdot p+2q\cdot pq\cdot k-q^2k\cdot p\right) \nonumber\\
&+\left(q^\mu k^\nu+q^\nu k^\mu\right)\left(p\cdot\overline{p}q\cdot k'-k'\cdot pq\cdot\overline{p}-k'\cdot\overline{p}q\cdot p\right) \nonumber\\
&+\left(q^\mu k'^\nu+q^\nu k'^\mu\right)\left(p\cdot\overline{p}q\cdot k-k\cdot\overline{p}q\cdot p-k\cdot pq\cdot\overline{p}\right) \nonumber\\
&+\left(k^\mu k'^\nu+k^\nu k'^\mu\right)\left(2q\cdot pq\cdot\overline{p}-q^2p\cdot\overline{p}\right) \nonumber\\
&+\left(q^\mu q^\nu+q^\nu q^\mu\right)\left(k'\cdot\overline{p}k\cdot p-k\cdot k'p\cdot\overline{p}+k'\cdot pk\cdot\overline{p}\right).
\end{align}
In the case of $V-A$ interaction, the amplitude squared is only $M_L$ which was given in \cite{Fujikawa:1971nx}.

Nextly, moving on to the explicit form of $M_R$, one can easily derive it by taking the charge conjugate of materials in the trace:
\begin{align}
M_R=& 4J_{\mu\nu}j^{\alpha\beta}\mathrm{Tr}\left[\Slash{p}W_\beta^\nu\Slash{\overline{p}}V_\alpha^\mu P_L\right] \nonumber\\
=& 4J_{\mu\nu}j^{\beta\alpha}\mathrm{Tr}\left[\Slash{p}W_\alpha^\mu\Slash{\overline{p}}V_\beta^\nu P_L\right].
\end{align}
One notices that the form in the last line is the same as that of $M_L$ except for the superscripts of the neutrino tensor $j^{\alpha\beta}$.
According to Eq.~\eqref{seq:neutrino-part}, the exchange of $\mu$ and $\nu$ in $j^{\mu\nu}$ clearly corresponds to the exchange of $k$ and $k'$.
Therefore, $M_R$ is obtained as 
\begin{align}
M_R=&\left(\text{$k\leftrightarrow k'$ exchange of $M_L$}\right).
\end{align}

At last, we present $M_{LR}$ and $M_{RL}$.
By using the explicit forms of $W_\alpha^\mu$ and $V_\beta^\nu$, one obtains 
$J_{\mu\nu}\mathrm{Tr}\left[W_\alpha^\mu V_\beta^\nu\gamma_5\right]=0$, which means $M_{LR}=M_{RL}$.
Then, the terms are given by
\begin{align}
M_{LR}=M_{RL}=&(-256J_{\mu\nu})\left[\frac{M_{LR1}^{\mu\nu}}{\left(q^2+2p\cdot q\right)^2}+\frac{M_{LR2}^{\mu\nu}}{\left(q^2+2\overline{p}\cdot q\right)^2}+\frac{M_{LR3}^{\mu\nu}}{\left(q^2+2p\cdot q\right)\left(q^2+2\overline{p}\cdot q\right)}\right], \\
M_{LR1}^{\mu\nu}=&\frac{m_\ell^2}{2}k\cdot k'\left\{q^2g^{\mu\nu}+2\left[p^\mu q^\nu+p^\nu q^\mu\right]+4p^\mu p^\nu\right\}, \\
M_{LR2}^{\mu\nu}=&\left(\text{$p\leftrightarrow\overline{p}$ exchange of $M_{LR1}^{\mu\nu}$}\right), \\
M_{LR3}^{\mu\nu}=& m_\ell^2\left[g^{\mu\nu}\left(2k\cdot qk'\cdot q-q^2k\cdot k'\right)-2k\cdot k'\left(p^\mu\overline{p}^\nu+p^\nu\overline{p}^\mu\right)\right. \nonumber\\
&-k\cdot k'\left(p^\mu q^\nu+p^\nu q^\mu+\overline{p}^\mu q^\nu+\overline{p}^\nu q^\mu\right) \nonumber\\
&\left.+q^2\left(k^\mu k'^\nu+k^\nu k'^\mu\right)-k'\cdot q\left(k^\mu q^\nu+k^\nu q^\mu\right)-k\cdot q\left(k'^\mu q^\nu+k'^\nu q^\mu\right)\right].
\end{align}

When the incident neutrino is an anti-neutrino, the result can be obtained by replacing $k\leftrightarrow k'$ in the above formulas.
Note that the obtained transition density is invariant under the simultaneous replacement of $k\leftrightarrow k'$ and $p\leftrightarrow\overline{p}$.
These facts imply that the roles of emitted charged leptons are completely exchanged in the anti-neutrino case.

%%%%%%%%%%%%%%%%%%%%%%%%%%%%%%%%%%%%%%%%%%%%%%%%%%%%%%%%%%
\section{Phase Space Integrals} \label{apdx:phase-space-int}
%%%%%%%%%%%%%%%%%%%%%%%%%%%%%%%%%%%%%%%%%%%%%%%%%%%%%%%%%%

We perform the Monte Carlo method in calculating the four-body phase space integral \cite{Czyz:1964zz, Brown:1973ih, Lovseth:1971vv}.
To achieve enough convergence of the integration, we choose suitable integral variables to flatten the integrand.
The phase space integrals for the four-body final state are
\begin{align}
d\Pi=\frac{d^3k'}{(2\pi)^32E_{k'}}\frac{d^3p}{(2\pi)^32E_{p}}\frac{d^3\overline{p}}{(2\pi)^32E_{\overline{p}}}\frac{d^3Q'}{(2\pi)^32E_{Q'}}\left(2\pi\right)^4\delta^{(4)}\left(k'+p+\overline{p}+Q'-k-Q\right).
\end{align}
Although the number of the integration variables are $3\times 4=12$, the net number is only eight because of the energy-momentum conservation.

In general, we can rewrite the phase space integral to
\begin{align}
\int d\Pi=&\int_{\underline{x}_0}^{\overline{x}_0}dx_0\int_{\underline{x}_1}^{\overline{x}_1}dx_1\int_{\underline{x}_2}^{\overline{x}_2}dx_2\int_{\underline{x}_3}^{\overline{x}_3}dx_3\int_{\underline{x}_4}^{\overline{x}_4}dx_4\int_{\underline{x}_5}^{\overline{x}_5}dx_5\int_{\underline{x}_6}^{\overline{x}_6}dx_6\int_{\underline{x}_7}^{\overline{x}_7}dx_7 \nonumber\\
&\times X(x_0,x_1,x_2,x_3,x_4,x_5,x_6,x_7),
\end{align}
where $X$ is an overall factor depending on the integral variables $x_i$ ($i=0,1,\cdots,7$).
The upper and lower limits of $x_i$ are represented by $\overline{x}_i$ and $\underline{x}_i$, respectively.
Since it is difficult to find the range of arbitrary integral variables, we have to choose a useful set of integral variables.
In our analysis, we use the following set of the integral variables $x_i$s:
\begin{align}
x_0=&\phi_{Q'}^{(A)}, \\
x_1=&\tau\equiv\int_{t}^{\infty}du\left\{F(u)\right\}^2, \label{eq:t2tau}\\
x_2=& s_{p\overline{p}k}\equiv\left(p+\overline{p}+k'\right)^2, \\
x_3=& s_{\overline{p}k}\equiv\left(\overline{p}+k'\right)^2, \\
x_4=& u_p\equiv\log\left(-q^2-2p\cdot q\right), \\
x_5=&\phi_p^{(B)}, \\
x_6=& u_{\overline{p}}\equiv\log\left(-q^2-2\overline{p}\cdot q\right), \\
x_7=&\phi_{\overline{p}}^{(C)}.
\end{align}
$\phi_{Q'}^{(A)}$, $\phi_{p}^{(B)}$, and $\phi_{\overline{p}}^{(C)}$ are rotation angles defined as follow:
$\phi_{Q'}^{(A)}$ is the rotation angle of $\bm{Q}'$ around $\bm{k}$ in the center-of-mass frame of $\bm{k}$ and $\bm{Q}$, which we call the frame $A$.
$\phi_{p}^{(B)}$ is the rotation angle of $\bm{p}$ around $\bm{q}$ in the frame where $\bm{p}+\overline{\bm{p}}+\bm{k}'=0$, which we call the frame $B$.
$\phi_{\overline{p}}^{(C)}$ is the rotation angle of $\overline{\bm{p}}$ around $\bm{q}$ in the frame where $\overline{\bm{p}}+\bm{k}'=0$, which we call the frame $C$.
This choice of variables is useful because all the three angles trivially run from $0$ to $2\pi$.
Here, we obtain the overall factor,
\begin{align}
X=\frac{\exp\left(u_{p}+u_{\overline{p}}\right)}{(4\pi)^8\left(s-M^2\right)\sqrt{s_{p\overline{p}k'}s_{\overline{p}k'}\left(t+\left|q_0^{\left(B\right)}\right|^2\right)\left(t+\left|q_0^{\left(C\right)}\right|^2\right)}\left\{F\left(t\right)\right\}^2}.
\end{align}

For the NTP processes, the differential cross sections have the rotational symmetry around the neutrino beam axis.
Then, the integral of $\int d\phi_{Q'}^{(A)}$ can be simply replaced by $2\pi$, and practically the other seven integral variables are relevant.

The variable $\tau$, defined by Eq.~(\ref{eq:t2tau}), runs over 
\begin{align}
\int_{t_\mathrm{max}}^{\infty}du\left\{F(u)\right\}^2<\tau<\int_{t_\mathrm{min}}^{\infty}du\left\{F(u)\right\}^2,
\end{align}
where $t_\mathrm{max}$ ($t_\mathrm{min}$) is the maximum (minimum) value of $t=-q^2$.
$t_\mathrm{max}$ and $t_\mathrm{min}$ are given by
\begin{align}
t_\mathrm{max}=&\frac{s}{2}\left(1-\frac{M^2}{s}\right)\left\{1-\frac{M^2}{s}+\sqrt{\lambda\left(1,\frac{M^2}{s},\frac{4m_\ell^2}{s}\right)}\right\}-2m_\ell^2\left(1+\frac{M^2}{s}\right),\\
t_\mathrm{min}=&\frac{4m_\ell^2}{t_\mathrm{max}}\frac{M^2}{s},
\end{align}
Since we do not have the analytic representation of $t$ as a function of $\tau$, we prepare the numerical correspondence table between $t$ and $\tau$ 
for the phase space integration.

The ranges of the rest variables are as follows:
\begin{align}
4m_\ell^2<s_{p\overline{p}k'}<&\frac{(s-M^2)\sqrt{t\left(t-4M^2\right)}-\left(s+M^2\right)t}{2M^2}, \\
m_\ell^2<s_{\overline{p}k'}<&\left(\sqrt{s_{p\overline{p}k'}}-m_\ell\right)^2,
\end{align}
\begin{align}
\underline{u}_p=&\log\left(t-\frac{s_{p\overline{p}k'}-s_{\overline{p}k'}+m_\ell^2}{\sqrt{s_{p\overline{p}k'}}}q_0^{\left(B\right)}-\sqrt{s_{p\overline{p}k'}\lambda\left(1,\frac{m_\ell^2}{s_{p\overline{p}k'}},\frac{s_{\overline{p}k'}}{s_{p\overline{p}k'}}\right)\left(t+\left|q_0^{\left(B\right)}\right|^2\right)}\right), \\
\overline{u}_p=&\log\left(t-\frac{s_{p\overline{p}k'}-s_{\overline{p}k'}+m_\ell^2}{\sqrt{s_{p\overline{p}k'}}}q_0^{\left(B\right)}+\sqrt{s_{p\overline{p}k'}\lambda\left(1,\frac{m_\ell^2}{s_{p\overline{p}k'}},\frac{s_{\overline{p}k'}}{s_{p\overline{p}k'}}\right)\left(t+\left|q_0^{\left(B\right)}\right|^2\right)}\right), \\
\underline{u}_{\overline{p}}=&\log\left(t-\frac{s_{\overline{p}k'}+m_\ell^2}{\sqrt{s_{\overline{p}k'}}}q_0^{\left(C\right)}-\frac{s_{\overline{p}k'}-m_\ell^2}{\sqrt{s_{\overline{p}k'}}}\sqrt{t+\left|q_0^{\left(C\right)}\right|^2}\right), \\
\overline{u}_{\overline{p}}=&\log\left(t-\frac{s_{\overline{p}k'}+m_\ell^2}{\sqrt{s_{\overline{p}k'}}}q_0^{\left(C\right)}+\frac{s_{\overline{p}k'}-m_\ell^2}{\sqrt{s_{\overline{p}k'}}}\sqrt{t+\left|q_0^{\left(C\right)}\right|^2}\right),
\end{align}
where $q_0^{(B)}$ and $q_0^{(C)}$ are the time component of $q$ in the frame $B$ and $C$, respectively.
In principle, we can derive the analytic formulas for $q_0^{(B)}$ and $q_0^{(C)}$, which are complicated a little.
However, we do not need the analytic formulas because we easily obtain the numerical values of $q_0^{(B)}$ and $q_0^{(C)}$ step-by-step in the Monte Carlo integration.
At each step, the upper and lower limit of $u_{p}$ are determined after $\tau$, $s_{p\overline{p}k'}$, and $s_{\overline{p}k'}$ are fixed.
Then, the upper and lower limit of $u_{\overline{p}}$ are determined after $u_p$ and $\phi_p^{(B)}$ are fixed in addition.

%%%%%%%%%%%%%%%%%%
%%% references %%%
%%%%%%%%%%%%%%%%%%
\bibliographystyle{apsrev}
\bibliography{biblio}

\end{document}